\documentclass[aps,prb,twocolumn,superscriptaddress,longbibliography]{revtex4-2}

\usepackage{graphicx}% Include figure files
\usepackage{dcolumn}% Align table columns on decimal point
\usepackage{bm}% bold math
\usepackage{subcaption}
\usepackage{braket}
\usepackage[table]{xcolor}
\usepackage{colortbl}
\usepackage{array}
\usepackage{hyperref}

\usepackage{tabularx}
\usepackage{tabularray}
\usepackage[normalem]{ulem}
\usepackage{bm}

\usepackage[utf8]{inputenc}
\usepackage{braket}
\usepackage{enumitem}
\usepackage{natbib}
\usepackage{float} 
\usepackage{etoolbox}
\usepackage{balance}
\usepackage[table]{xcolor}
\usepackage{colortbl}
\usepackage{array} 
\usepackage{xcolor}

\usepackage{tabularx}
\usepackage{tabularray}
\usepackage{bm}

\usepackage[utf8]{inputenc} % not needed with modern engines, but harmless
\usepackage[T1]{fontenc}
\usepackage{amsmath,amssymb,amsfonts}
\usepackage{graphicx}
\usepackage{dcolumn}   % Align table columns on decimal point
\usepackage{bm}        % Bold math
\usepackage{fancyhdr}
\usepackage{hyperref}

\hypersetup{
  colorlinks = true,
  linkcolor  = blue,
  citecolor  = blue,
  urlcolor   = blue,
  breaklinks = true
}

\begin{document}

\preprint{APS/14.03.2006}

\title{\textbf{From Approximate Floquet Engineering to Full Floquet Theory: Coherent Control of Chiral Spin Systems in Spintronics} 
}% 

\author{Andrea Simion}
\affiliation{
Faculty of Physics, Babeș Bolyai University, 400084 Cluj-Napoca, Romania
}
\affiliation{
National Institute for Research and Development of Isotopic and Molecular Technologies, 400296 Cluj-Napoca, Romania
}

\author{Claudiu Filip}
\affiliation{
National Institute for Research and Development of Isotopic and Molecular Technologies, 400296 Cluj-Napoca, Romania
}

\author{Coriolan V. Tiușan}
\email{Corresponding author: coriolan.tiusan@ubbcluj.ro} 
\affiliation{
Faculty of Physics, Babeș Bolyai University, 400084, Cluj-Napoca, Romania
}
\affiliation{
National Center of Scientific Research, 54000 Nancy, France
}


\begin{abstract}
Coherent control of interacting spin systems under time-periodic driving is a central challenge in spin-based quantum technologies. Here we demonstrate the applicability of a full Floquet-space formalism, adapted from Nuclear Magnetic Resonance (NMR) methodologies, to model the dynamics of driven coupled electron spins in the presence of a static magnetic field $B_{0}$ and a transverse oscillating field $B_{1}$. The framework explicitly includes isotropic exchange coupling \textit{J} and the chiral Dzyaloshinskii–Moriya antisymmetric exchange interaction (DMI), and its numerical convergence is systematically validated with respect to Fourier-space truncation. In the non-interacting limit, the expected driven-spin dynamics is recovered, with the oscillation periodicity governed by $B_{1}$. Exchange coupling alone does not modify the collective spin expectation values under the chosen initial condition, consistent with symmetry considerations. In contrast, increasing DMI generates a finite  $\langle \hat{S}_{y} \rangle$ component, suppresses $\langle \hat{S}_{z}\rangle$, and produces tilted, elliptical Bloch-sphere trajectories, reflecting the emergence of chiral spin–spin correlations. These effects are pronounced for open boundary conditions, while remaining nearly negligible in the periodic boundary case. When exchange coupling and DMI coexist, the dynamics becomes strongly perturbed and multi-frequency in nature. Together, these results demonstrate that full Floquet-space modeling provides a robust and predictive framework for analyzing and engineering coherent dynamics in driven interacting spin systems beyond simple coherent-rotation regimes.
\end{abstract}

\maketitle

\section{Introduction}
\label{sec:intro}
The coherent manipulation of interacting spins under time-periodic driving is central to magnetic resonance, quantum information processing, and spintronics. In solid-state nuclear magnetic resonance (NMR), radio-frequency irradiation is routinely used to engineer effective Hamiltonians, suppress unwanted couplings \cite{d1, d2, d3}, and selectively restore interactions \cite{r1, r2, r3}. Related control concepts are increasingly important in spintronic and quantum-spin platforms, where periodic driving can manipulate collective excitations, spin correlations, and chiral magnetic states.

Coherent spin control has been demonstrated across a broad range of physical systems \cite{n0, n1, n2, n3, n4, n5, n6, n7, n8, n9, n10, n11}. Electron-spin resonance enabled Rabi oscillations and coherent gate operations in semiconductor quantum dots \cite{int1}, while electric-dipole spin resonance provided electrically driven rotations mediated by spin–orbit coupling \cite{int2}. In magnetic heterostructures, spin-transfer and spin–orbit torques have enabled magnetization switching, coherent excitation, and microwave generation \cite{int3, int7, int4}. Coherent dynamics has also been extended to the terahertz regime in antiferromagnetic materials \cite{int5, int6}. Of particular interest are chiral magnetic systems governed by the Dzyaloshinskii–Moriya interaction (DMI), which supports non-collinear spin configurations and topological textures such as skyrmions, with potential applications in classical, neuromorphic, and quantum technologies \cite{int9, int10, int11}.

Floquet engineering provides a general framework for controlling such systems through periodic driving \cite{f1, f2, f3, f4, f7, int8, f6, f5, f8}. A time-periodic Hamiltonian can be represented in terms of Fourier components and mapped onto a time-independent problem in an extended Hilbert–Floquet, or Sambe, space. In many applications, however, the driven system is replaced by an effective static Hamiltonian obtained from a truncated high-frequency Floquet–Magnus expansion. At first order, this approach retains only the leading correction proportional to the inverse driving frequency. It is therefore controlled only when the driving frequency is sufficiently large compared with the internal interaction scales. This condition is not necessarily satisfied in interacting spin systems, where the driving, isotropic exchange, and chiral interactions may have comparable characteristic energies. In this regime, neglected higher-order contributions, multiphoton processes, quasienergy avoided crossings, and resonance effects can substantially modify the dynamics. A low-order effective Hamiltonian may consequently predict regular, nearly single-frequency precession while missing amplitude modulations, additional frequencies, and complex magnetization trajectories. Establishing the range of validity of approximate Floquet engineering therefore requires direct comparison with a systematically converged treatment of the full time-periodic Hamiltonian.

Full operator-based Floquet methods have long been developed in magnetic resonance \cite{floq6, floq7, floq5, floq4, floq3, floq2, floq1}. Rather than truncating an inverse-frequency expansion, these methods construct the Floquet Hamiltonian explicitly in an extended Fourier space. Its diagonalization provides the quasienergies and Floquet modes, while convergence is controlled by increasing the number of retained Fourier sectors. This approach is computationally more demanding than a low-order effective Hamiltonian, but it remains applicable beyond the strict high-frequency regime and retains the higher-order and resonant processes responsible for multifrequency dynamics. These effects are especially relevant in the presence of DMI. Unlike isotropic exchange, DMI mixes spin components, generates chiral correlations, and transfers polarization into non-collinear and correlated spin degrees of freedom. Its influence also depends strongly on system geometry. Periodic boundary conditions preserve translational equivalence among sites, whereas open boundaries break this equivalence and permit spatially nonuniform spin responses \cite{Tiusan2025}. In short spin chains, where finite-size and boundary effects are not averaged out, this distinction can qualitatively reshape the driven dynamics. Such systems therefore provide controlled minimal models for determining when an approximate Floquet description remains adequate and when a full Floquet-space treatment becomes necessary. 

In this work, we present, to our knowledge, the first systematically converged full Floquet-space treatment of coherent control in interacting chiral electron-spin systems. The formalism is adapted from NMR and applied to spins subject to a static magnetic field, a transverse periodic magnetic field, isotropic exchange coupling, and DMI. Its convergence is established with respect to the Fourier-space truncation order. We then directly compare the resulting dynamics with those obtained from the first-order Floquet–Magnus effective Hamiltonian. When the driving and interaction scales are comparable, or when the boundaries are open, the approximate theory produces artificially regular trajectories and fails to reproduce the amplitude modulation, additional frequency components, and complex magnetization paths captured by the full Floquet treatment. We first use a two-spin system to benchmark the method and isolate the effects of the different interaction terms. The non-interacting limit recovers the expected driven-spin dynamics, while isotropic exchange alone preserves the collective-spin expectation values for the chosen initial state. In contrast, DMI generates a finite transverse magnetization, suppresses the longitudinal component, and transforms regular Bloch-sphere rotations into tilted and elliptical trajectories. When exchange and DMI coexist, the dynamics becomes strongly perturbed and multifrequency. We then extend the treatment to three-spin systems with open and periodic boundary conditions. Open chains exhibit pronounced interaction-induced distortions, whereas periodic closure partially restores coherent collective motion. Finally, we construct control landscapes over the static- and driving-field parameter space to identify conditions that generate targeted NMR-like spin rotations despite the competing interactions. These results establish the full Floquet formalism as a predictive framework for coherent control beyond perturbative Floquet engineering and clarify the regimes in which first-order Floquet–Magnus theory is insufficient for driven chiral spintronic and quantum-spin systems.

\section{Theory}
\begin{figure*}
 \centering
        \includegraphics[width=0.8\textwidth]{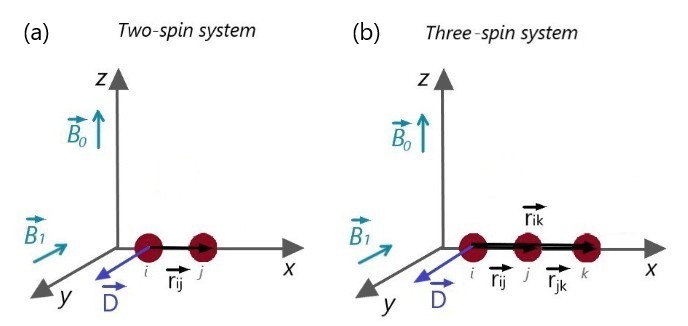}
\caption{\label{fig:wide}Schematic representation of a two-spin (a) and three-spin (b) Ising chain, showing the orientations of the magnetic fields  $\vec{B}_{0}$, $\vec{B}_{1}$, and the DMI vector $\vec{D}$}.
\label{fig:1}
\end{figure*}

For the beginning, we consider the simplest case of a two-spin system Ising chain model, which is schematically represented in Figure 1 (a). The system is subjected to a static magnetic field ($B_{0}$), an oscillating magnetic field ($B_{1}$),  an exchange interaction (J), and a DMI interaction (D,  oriented along the Oy axis). Coonsidering that the spins are oriented along the Oz axis,  the initial state of the system is given by:
\begin{equation}
    \hat{\rho}(0)=\hat{S}_{1z}+\hat{S}_{2z}
\end{equation}
where $\hat{\rho}_{0}$ is the initial density operator, while $\hat{S}_{1z}$ and $\hat{S}_{2z}$ are the \textit{z}-components of the two spins. This initial state would naturally correspond to a groundstate of the system defined by $\vec{B}_{0} \ne 0$ in the absence of DMI and rotating field ($\vec{B}_{1}=0$) .

Taking into account that the DMI term is perpendicular to the Ising chain, i.e., DMI is on the Oy axis, the total Hamiltonian in the laboratory frame, describing the energy of the spin system, is given by:

\begin{equation}
\begin{split}
    \hat{H}
=&
-J \left( \hat{S}_{1x}\hat{S}_{2x} + \hat{S}_{1y}\hat{S}_{2y} + \hat{S}_{1z}\hat{S}_{2z} \right) \\
+ &D \left( \hat{S}_{1z}\hat{S}_{2x} - \hat{S}_{1x}\hat{S}_{2z} \right)
+ \omega_0 \left( \hat{S}_{1z} + \hat{S}_{2z} \right)
\\+& \omega_1 \left( \hat{S}_{1y} + \hat{S}_{2y} \right)\cos(\omega_{0} t)
\end{split}
\end{equation}
where $\omega_{0}$ and $\omega_{1}$ are the amplitudes of the two magnetic fields, defined as $\omega_{0}=\gamma B_{0}$ and $\omega_{1}=\gamma B_{1}$, $\gamma$ being the gyromagnetic ratio of the electron. Note that the minus sign of the exchange term indicates that $J>0$ would correspond to a ferromagnetic-type interaction between adjacent spins and $J<0$ to an antiferromagnetic one.

To obtain the Hamiltonian in the rotating frame (with the frequency $\omega_{0}$), one performs the following transformation:
\begin{equation}
    \hat{H}_{rot} = e^{-i \omega_{0}\hat{S}_{z}t} \hat{H}e^{i \omega_{0}\hat{S}_{z}t} -\omega_0 \left( \hat{S}_{1z} + \hat{S}_{2z} \right)
\end{equation}
After some trivial calculations, the Hamiltonian in the rotating frame can be written as:
\begin{equation}
\begin{split}
    \hat{H}
=&
-J \left( \hat{S}_{1x}\hat{S}_{2x} + \hat{S}_{1y}\hat{S}_{2y} + \hat{S}_{1z}\hat{S}_{2z} \right) +\omega_1 \left( \hat{S}_{1y} + \hat{S}_{2y} \right) \\
+& \frac{D}{2} \left[ \left( \hat{S}_{1z}\hat{S}_{2+} - \hat{S}_{1+}\hat{S}_{2z} \right)e^{i\omega_{0}t}
+ \left( \hat{S}_{1z}\hat{S}_{2-} - \hat{S}_{1-}\hat{S}_{2z} \right)e^{-i\omega_{0}t}\right]
\end{split}
\end{equation}
where the first two terms represent the isotropic terms, while the third one is modulated in time with the $\omega_{0}$ frequency (or, driving frequency in the rotating frame). To transform this time-dependent Hamiltonian into a time-independent Hamiltonian, one applies the General Floquet Theory. The Schrödinger equation is given by:
\begin{equation}
    i\,\frac{d}{dt}\,|\psi(t)\rangle =
\hat{H}(t)\,|\psi(t)\rangle
\end{equation}
The General Floquet Theory say that all solution can be written as:
\begin{equation}
    |\psi_\alpha(t)\rangle
=
\mathrm{e}^{-i \varepsilon_\alpha t}
\,|u_\alpha(t)\rangle
\end{equation}
where $\varepsilon_\alpha$ are the quasienergies and $\,|u_\alpha(t)\rangle$ are periodic states, i.e. $\,|u_\alpha(t+T)\rangle=\,|u_\alpha(t)\rangle$. From eqs. (5) and (6), one obtain:
\begin{equation}
    \left(
\hat{H}(t)
- i\,\frac{d}{dt}
\right)
|u_\alpha(t)\rangle
=
\varepsilon_\alpha
|u_\alpha(t)\rangle
\end{equation}
which represent an eigenvalue problem for the operator:
\begin{equation}
    \hat{H}_{F}=\hat{H}(t)
- i\,\frac{d}{dt}
\end{equation}
where $\hat{H_{F}}$ is the Floquet Hamiltonian, acting on time-periodic states $|u_\alpha(t)\rangle$. Then, $\hat{H_{F}}$ is a time-independent Hamiltonian which acts in a larger space, i.e., Hilbert space $\otimes$ T-periodic functions. For any periodic operator $\hat{H}(t)$, one can expand this in a Fourier series:
\begin{equation}
\begin{split}
    \hat{H}(t) =&
\sum_{n=-\infty}^{\infty}
\hat{H}_{n} \,
\mathrm{e}^{i n \omega t}\\
\hat{H}_{n}
=&
\frac{1}{T}
\int_{0}^{T}
dt \,
\hat{H}(t)\,
\mathrm{e}^{-i n \omega t}
\end{split}
\end{equation}
In our Hamiltonian $\hat{H}_{rot}$, the first two terms represent the $n=0$ component (they are time-independent terms), while the last one has two modulation frequencies, corresponding to the $n = \pm 1$ components. Therefore, one obtains the $0$, $+1$, and $-1$ Fourier components of the time-periodic Hamiltonian:
\begin{equation}
\begin{split}    
\hat{H}_0
&=
-J \left(
\hat{S}_{1x}\hat{S}_{2x}
+ \hat{S}_{1y}\hat{S}_{2y}
+ \hat{S}_{1z}\hat{S}_{2z}
\right)
+ \omega_1 \left( \hat{S}_{1y} + \hat{S}_{2y} \right) \\
\hat{H}_{+1}
&=
\frac{D}{2} \left( \hat{S}_{1z}\hat{S}_{2+} - \hat{S}_{1+}\hat{S}_{2z} \right) \\
\hat{H}_{-1}
&=
\frac{D}{2}\left( \hat{S}_{1z}\hat{S}_{2-} - \hat{S}_{1-}\hat{S}_{2z} \right)
\end{split}
\end{equation}
The total Hamiltonian in the Floquet space is given by:
\begin{equation}
    \hat{H}_{F} = \hat{I}_{F} \otimes \hat{H}_{0} + \hat{K}_{+1} \otimes \hat{H}_{+1} + \hat{K}_{-1} \otimes \hat{H}_{-1} + \omega_{0}\hat{N} \otimes \hat{I}_{spins}
    \tag{5.2}
\end{equation}
where  $\hat{I}_{F}$ is the identity operator in Floquet space, $\hat{K}_{+1}$ and $\hat{K}_{-1}$ are the ladder and the shift operators, $\hat{N}$ is the Floquet number operator, and $\hat{I}_{spins}$ is the identity operator in spin Hilbert space. These operators are given by:
\begin{equation}
    \begin{split}
        \hat{I}_{F} =&
\sum_{m=-\infty}^{+\infty}
|m\rangle \langle m| \\
\hat{K}_{+1}
=&
\sum_{m=-\infty}^{\infty}
|m+1\rangle \langle m| \\
\hat{K}_{-1}
=&
\sum_{m=-\infty}^{\infty}
|m-1\rangle \langle m|\\
\hat{N}
=&
\sum_{m=-\infty}^{\infty}
m \, |m\rangle \langle m|\\
\hat{I}_{\mathrm{spins}}
=&
\sum_{\alpha=1}^{d_{\mathrm{spin}}}
|\alpha\rangle \langle \alpha|
    \end{split}
\end{equation}
where $d_{spin}$ is the dimension of the spin Hilbert space. To obtain the evolution of the initial density operator in time, one applies the Liouville von Neumann equation:
\begin{equation}
\frac{d\hat{\rho}(t)}{dt}
=
-\,i \left[ \hat{H}_F , \hat{\rho}(t) \right] 
\end{equation}
while the expected values of the spin system components are obtained by:

\begin{equation}
\begin{split}
    \langle\hat{S}_{x}(t)\rangle =& Tr\{\hat{S}_{x} \hat{\rho}(t) \}\\
    \langle\hat{S}_{y}(t)\rangle =& Tr\{\hat{S}_{y} \hat{\rho}(t) \}\\
    \langle\hat{S}_{z}(t)\rangle =& Tr\{\hat{S}_{z} \hat{\rho}(t) \}
\end{split}
\end{equation}

 For the three-spin system (Figure 1 (b)), one can divide the discussion into two casess: (i) Ising chain with open boundary condition (OBC), when spin 1 is coupled to spin 2 and spin 2 is coupled to spin 3, but there is no coupling between spins 1 and 3, and (ii) Ising chain with periodic boundary conditions (PBC), when there is coupling also between the first and the last spin (1 and 3). Using the same formalism, one obtains the $n=0$ and $ n= \pm 1$ Fourier components of the Hamiltonian. For open boundary conditions, one has:
  \begin{equation}
     \begin{split}    
\hat{H}_0
=&
-J \left(
\hat{S}_{1x}\hat{S}_{2x}
+ \hat{S}_{1y}\hat{S}_{2y}
+ \hat{S}_{1z}\hat{S}_{2z}
\right) \\
-& J \left(
\hat{S}_{2x}\hat{S}_{3x}
+ \hat{S}_{2y}\hat{S}_{3y}
+ \hat{S}_{2z}\hat{S}_{3z}
\right)\\
+& \omega_1 \left( \hat{S}_{1y} + \hat{S}_{2y} + \hat{S}_{3y} \right) \\
\hat{H}_{+1}
&=
\frac{D}{2} \left( \hat{S}_{1z}\hat{S}_{2+} - \hat{S}_{1+}\hat{S}_{2z} \right) + \frac{D}{2}\left( \hat{S}_{2z}\hat{S}_{3+} - \hat{S}_{2+}\hat{S}_{3z} \right) \\
\hat{H}_{-1}
&=
\frac{D}{2} \left( \hat{S}_{1z}\hat{S}_{2-} - \hat{S}_{1-}\hat{S}_{2z} \right)+\frac{D}{2}\left( \hat{S}_{2z}\hat{S}_{3-} - \hat{S}_{2-}\hat{S}_{3z} \right)
\end{split}
 \end{equation}
 while for periodic boundary conditions, the Fourier components are given by:
 \begin{equation}
     \begin{split}    
\hat{H}_0
=&
-J \left(
\hat{S}_{1x}\hat{S}_{2x}
+ \hat{S}_{1y}\hat{S}_{2y}
+ \hat{S}_{1z}\hat{S}_{2z}
\right)\\
-& J \left(
\hat{S}_{2x}\hat{S}_{3x}
+ \hat{S}_{2y}\hat{S}_{3y}
+ \hat{S}_{2z}\hat{S}_{3z}
\right)\\
-& J \left(
\hat{S}_{3x}\hat{S}_{1x}
+ \hat{S}_{3y}\hat{S}_{1y}
+ \hat{S}_{3z}\hat{S}_{1z}
\right) \\
+&\omega_1 \left( \hat{S}_{1y} + \hat{S}_{2y} + \hat{S}_{3y} \right) \\
\hat{H}_{+1}
=&
\frac{D}{2} \left( \hat{S}_{1z}\hat{S}_{2+} - \hat{S}_{1+}\hat{S}_{2z} \right) +\frac{D}{2} \left( \hat{S}_{2z}\hat{S}_{3+} - \hat{S}_{2+}\hat{S}_{3z} \right) \\
+&\frac{D}{2} \left( \hat{S}_{3z}\hat{S}_{1+} - \hat{S}_{3+}\hat{S}_{1z} \right) \\
\hat{H}_{-1}
=&
\frac{D}{2} \left( \hat{S}_{1z}\hat{S}_{2-} - \hat{S}_{1-}\hat{S}_{2z} \right)+\frac{D}{2} \left( \hat{S}_{2z}\hat{S}_{3-} - \hat{S}_{2-}\hat{S}_{3z} \right)\\
+&\frac{D}{2} \left( \hat{S}_{3z}\hat{S}_{1-} - \hat{S}_{3-}\hat{S}_{1z} \right)
\end{split}
 \end{equation}
The Floquet Hamiltonians constructed in this extended Hilbert–Floquet space are formally infinite-dimensional, as it includes an infinite number of Fourier (Floquet) modes associated with the time-periodic driving. In practical calculations, this space must be truncated to a finite number of modes while preserving the accuracy of the dynamics. Therefore, one needs to perform a controlled truncation by restricting the Floquet index to $m\in [-p, +p]$, resulting in a finite Floquet space of dimension $2p+1$. The full Hilbert–Floquet Hamiltonian then has dimension $d_{spin}$x$(2p+1)$, where $d_{spin}$ is the dimension of the spin Hilbert space. The truncation order $p$ is chosen based on physical and numerical convergence criteria rather than on a priori assumptions. Physically, higher-order Floquet modes correspond to processes involving the absorption or emission of multiple quanta of the driving frequency $\omega$ whose contribution rapidly decreases as the coupling strength between Floquet sectors becomes small compared to $m\omega$. In particular, for a Hamiltonian with finite Fourier components $\hat{H}_{n}$, the dominant couplings connect only neighboring or low-order Floquet blocks, and higher-order blocks become energetically detuned by multiples of $\omega$. Numerically, convergence is assessed by systematically increasing the truncation order $p$ and monitoring the stability of relevant observables, including quasienergies, spin expectation values, and Bloch-sphere trajectories. The truncation is considered converged when further increasing $p$ does not produce measurable changes in these quantities within the desired accuracy. This convergence-controlled truncation procedure follows the established operator-based Floquet methodology developed in solid-state NMR and ensures that the resulting Floquet Hamiltonian provides an accurate and predictive description of the driven interacting spin dynamics beyond perturbative or single-frequency approximations.

In contrast, in the Floquet-engineering approach based on the first-order Floquet–Magnus expansion, instead of solving eq. (7) in the full Sambe space and obtaining Floquet modes, quasienergies, exact avoided crossings, and exact multiphoton resonances, the periodically driven system is replaced by a static effective Hamiltonian:
\begin{equation}
\hat{H}_{\mathrm{eff}}
=
\hat{H}_0
+
\hat{H}^{(1)}
+
\hat{H}^{(2)}
+\cdots
\end{equation}
where only the first correction
\(
O(1/\Omega)
\) is kept. Therefore, in this approximation, the analysis starts from a time-dependent Hamiltonian, and immediately, a high-frequency expansion is performed, keeping only the first term in \(1/\Omega\):
\begin{equation}
\hat{H}_{\mathrm{eff}}
=
\hat{H}_0
+
\frac{1}{\Omega}
\sum_{m>0}
\frac{\left[\hat{H}_{-m},\hat{H}_m\right]}{m}
\end{equation}
As a result, no Floquet Hamiltonian is constructed, no Floquet matrix is diagonalized, no quasienergy spectrum is calculated, no Floquet modes are obtained, and the validity of the formalism requires that \(\Omega\) (the driving)  to be larger than the internal interaction scale \cite{int8}.

For the two-spin system, the calculated effective Hamiltonian in the approximate Floquet formalism is given by:
\begin{equation}
\hat{H}_{\mathrm{eff}}^{(1)}
=
\hat{H}_0
+
\frac{\left[\hat{H}_{-1},\hat{H}_{+1}\right]}{\omega_0}
\end{equation}
Using the Fourier components
\(\hat{H}_0\), \(\hat{H}_{+1}\), and \(\hat{H}_{-1}\) given in eq. (10), one obtain the commutator between the \(\hat{H}_{-1}\) and \(\hat{H}_{+1}\) components for spin-1/2:
\begin{equation}
    \begin{split}
    \left[
\hat{H}_{-1},
\hat{H}_{+1}
\right]=-\frac{D^2}{8}\left(\hat{S}_{1z}+\hat{S}_{2z}\right)
\end{split}
\end{equation}
Therefore, the approximate Floquet Hamiltonian becomes:
\begin{equation}
\begin{split}
\hat{H}_{\mathrm{eff}}^{(1)}
&=
-J
\left(
\hat{S}_{1x}\hat{S}_{2x}
+
\hat{S}_{1y}\hat{S}_{2y}
+
\hat{S}_{1z}\hat{S}_{2z}
\right)
+
\omega_1
\left(
\hat{S}_{1y}
+
\hat{S}_{2y}
\right) \\&
-\frac{D^2}{8 \omega_0}\left(\hat{S}_{1z}+\hat{S}_{2z}\right)
\end{split}
\end{equation}
Similarly, the effective Hamiltonian for a three-spin system with open boundary conditions using the approximate Floquet theory is given by:
\begin{equation}
    \begin{split}
\hat{H}_{\mathrm{eff}}^{(1)}
&=
-J
\left(
\hat{S}_{1x}\hat{S}_{2x}
+
\hat{S}_{1y}\hat{S}_{2y}
+
\hat{S}_{1z}\hat{S}_{2z}
\right)\\&
-J
\left(
\hat{S}_{2x}\hat{S}_{3x}
+
\hat{S}_{2y}\hat{S}_{3y}
+
\hat{S}_{2z}\hat{S}_{3z}
\right)
\\&+
\omega_1
\left(
\hat{S}_{1y}
+
\hat{S}_{2y}
+
\hat{S}_{3y}
\right) \\&
+
\frac{D^2}{4\omega_0}
\Big[
-\frac{1}{2}\hat{S}_{1z}
-
\hat{S}_{2z}
-
\frac{1}{2}\hat{S}_{3z}
+
4\hat{S}_{1z}\hat{S}_{2z}\hat{S}_{3z}
\\
&
+
\hat{S}_{1z}
\left(
\hat{S}_{2+}\hat{S}_{3-}
+
\hat{S}_{2-}\hat{S}_{3+}
\right)\\&
+
\left(
\hat{S}_{1+}\hat{S}_{2-}
+
\hat{S}_{1-}\hat{S}_{2+}
\right)
\hat{S}_{3z}
\Big]
\end{split}
\end{equation}
while for a three-spin system with periodic boundary conditions one obtains:
\begin{equation}
    \begin{split}
\hat{H}_{\mathrm{eff}}^{(1)}
=
&-J
\left(
\hat{S}_{1x}\hat{S}_{2x}
+
\hat{S}_{1y}\hat{S}_{2y}
+
\hat{S}_{1z}\hat{S}_{2z}
\right)\\&
-J
\left(
\hat{S}_{2x}\hat{S}_{3x}
+
\hat{S}_{2y}\hat{S}_{3y}
+
\hat{S}_{2z}\hat{S}_{3z}
\right)
\\
&
-J
\left(
\hat{S}_{3x}\hat{S}_{1x}
+
\hat{S}_{3y}\hat{S}_{1y}
+
\hat{S}_{3z}\hat{S}_{1z}
\right)\\&
+
\omega_1
\left(
\hat{S}_{1y}
+
\hat{S}_{2y}
+
\hat{S}_{3y}
\right)\\
&+
\frac{D^2}{4\omega_0}
\Big[
-\hat{S}_{1z}
-\hat{S}_{2z}
-\hat{S}_{3z}
+
12\hat{S}_{1z}\hat{S}_{2z}\hat{S}_{3z}\\&
+
2\hat{S}_{1z}
\left(
\hat{S}_{2+}\hat{S}_{3-}
+
\hat{S}_{2-}\hat{S}_{3+}
\right)
\\
&
+
2\hat{S}_{2z}
\left(
\hat{S}_{3+}\hat{S}_{1-}
+
\hat{S}_{3-}\hat{S}_{1+}
\right)\\&
+
2\hat{S}_{3z}
\left(
\hat{S}_{1+}\hat{S}_{2-}
+
\hat{S}_{1-}\hat{S}_{2+}
\right)
\Big]
\end{split}
\end{equation}
The analytical expressions obtained using the approximate Floquet theory provide a transparent first-order description of the effective interactions. However, because they neglect higher-order and resonant processes, their validity must be assessed by direct comparison with the systematically converged full Floquet treatment presented above.

\section{Materials and methods}
The theoretical framework developed in the previous section was compared to the appoximate Floquet framework and implemented numerically to analyze the coherent dynamics of driven interacting spin systems. Three cases were considered: (i) two-spin system with open boundary conditions (OBC), (ii) three-spin system with open boundary conditions (OBC), and (iii) three-spin system with periodic boundary conditions (PBC). Note that for a two-spin system, when periodic boundary conditions are connsidered a simple analytical calculation illustrates that the DMI has no effect, the respective interaction terms reciprocally canceling when closing the interaction loop $1 \rightarrow 2 \rightarrow 1$. These were subjected to a static magnetic field $B_{0}$ and an oscillating magnetic field $B_{1}$. Four representative regimes were considered: (i) the non-interacting case, where only the two magnetic fields were considered, (ii) the exchange-only case, where the exchange interaction (\textit{J}) was additionally included, (iii) the DMI-only regime, where, besides the magnetic fields, the DMI interaction was considered, neglecting the exchange interaction, and (iv) a combined exchange interaction and
DMI regime. The orientation of the magnetic fields and DMI was that presented in Figure 1: $B_{0}$ on the Oz-axis, and $B_{1}$ and DMI on the Oy-axis, while the initial direction of the spins was considered to be on the Oz-axis, for all the analyzed cases.

The time-dependent spin Hamiltonian was treated using an operator-based Floquet formalism adapted from nuclear magnetic resonance methodologies, allowing the periodically driven problem to be mapped onto a time-independent representation in an extended Hilbert–Floquet space. This approach enables a systematic treatment of multi-frequency dynamics and strong spin–spin interactions without relying on perturbative assumptions. The Floquet Hamiltonian was constructed by combining the spin Hilbert space with a Fourier space associated with the periodic driving. Since the Floquet representation is formally infinite-dimensional, a controlled truncation of the Fourier space was performed. The truncation order was not fixed a priori but was determined through explicit convergence tests. Physical observables, such as spin expectation values and magnetization trajectories, were monitored as the number of retained Fourier modes was progressively increased. The truncation was considered sufficient when further enlargement of the Floquet space did not produce measurable changes in these quantities and will be discussed in the results section.

Spin dynamics was evaluated by diagonalizing the truncated Floquet Hamiltonian in the rotating frame, and reconstructing time-dependent expectation values from the resulting Floquet eigenstates. Collective spin components were computed and analyzed to characterize coherent motion and correlation effects. The results were visualized using time-domain trajectories and Bloch-sphere representations, providing direct insight into deviations from simple coherent rotations.

The timescale shown in the figures is expressed in seconds, assuming that all Hamiltonian parameters are specified in frequency units (Hz). Within this convention, the interaction strengths, static field, and driving amplitudes directly define the natural dynamical timescale of the system, with time measured in inverse frequency units. Consequently, the evolution times presented here should be interpreted as normalized dynamical times determined by the chosen energy scales. This representation preserves the generality of the analysis, as the results can be straightforwardly rescaled to any physical system by adopting the appropriate interaction magnitudes.

All numerical calculations were performed using custom-developed routines in MATLAB. The methodology is general and can be readily extended to larger spin systems or more complex interaction networks, providing a robust framework for studying coherent control in driven interacting spin systems.

\section{Results and discussions}
\subsection{Floquet-space convergence and truncation order for the full Floquet theoretical formalism}
Before analyzing the driven spin dynamics, it is essential to establish the numerical convergence of the Floquet-space representation. Since the Floquet Hamiltonian is formally infinite-dimensional, a controlled truncation of the Fourier (Floquet) space is required. In this subsection, we systematically assess the impact of the truncation order on the calculated dynamics by increasing the number of retained Floquet modes and monitoring the resulting spin expectation values and magnetization trajectories. This procedure allows us to identify the minimal Floquet-space size required to accurately capture the interplay between periodic driving, exchange coupling, and chiral interactions, and to ensure that all subsequent results are free from truncation artifacts. 

Figure 2 illustrates the time evolution of the spin expectation values $\langle \hat{S}_{x}(t)\rangle$, $\langle \hat{S}_{y}(t)\rangle$, and $\langle\hat{S}_{z}(t)\rangle$, over 100 s for a driven two-spin system, together with the corresponding total magnetization trajectory on the Bloch sphere at the final time. The calculations are performed for fixed parameters, $B_{0}=1$, $B_{1}=0.5$, $J=1$, and $DMI=1$, while systematically increasing the Floquet-space truncation order, corresponding to Floquet indices $m\in[-1, 1]$ to $m\in[-6, 6]$ in panels (a)–(f).  The $J$ and $DMI$ values correspond to a strong symmetric and antisymmetric exchange interaction limit. As mentioned in the section Materials and methods, the field and exchange constants are in frequency units - Hz. 

For the smallest truncation ($m\in[-1, 1]$, panel a), the time traces already capture the qualitative oscillatory behavior induced by the periodic driving, but noticeable distortions are present. The oscillation amplitudes and relative phases between spin components are not yet stable, and the Bloch-sphere trajectory deviates from a smooth closed curve, indicating that relevant frequency components of the dynamics are missing at this truncation level. As the Floquet space is enlarged to $m\in[-2, 2]$ and $m\in[-3, 3]$ (panels b and c), the dynamics become progressively more regular. The time-domain signals converge toward stable oscillation envelopes, and the Bloch-sphere trajectories evolve into well-defined tilted elliptical orbits. This behavior reflects the inclusion of additional Fourier components required to correctly capture the interplay between coherent driving, isotropic exchange, and chiral Dzyaloshinskii–Moriya interactions, which together generate multi-frequency dynamics beyond simple single-frequency precession. For truncation orders $|m| \geq 3$ (panels c–f), no significant qualitative or quantitative changes are observed in either the spin expectation values or the Bloch-sphere trajectories. The oscillation amplitudes, phase relations, and the geometry of the magnetization orbit remain essentially unchanged upon further increasing the Floquet index. This demonstrates that the Floquet Hamiltonian has reached numerical convergence at $m\in[-3, 3]$ for the chosen parameter regime, even in the presence of strong interactions and chiral coupling. Physically, the tilted and elliptical Bloch-sphere trajectories observed in the converged regime are a direct consequence of the DMI, which mixes spin components and redistributes magnetization into correlated two-spin channels. Unlike the exchange interaction alone, which preserves simple collective rotations under the chosen initial conditions, the DMI introduces additional transverse components and breaks the symmetry of the motion, leading to non-circular orbits and multi-frequency behavior. Capturing these effects requires retaining a sufficient number of Floquet modes, as demonstrated by the clear convergence trend in Fig. 2. Similar simulations were performed for the three-spin systems, with open and periodic boundary conditions, where the numerical convergence was obtained for $m\in[-4, 4]$ (see Appendix A,  Figures $A1$ and $A2$). Therefore, for the next simulations, a truncation order with a Floquet index $m\in[-3, 3]$ was used for the two-spin system, and $m\in[-4, 4]$ for the three-spin systems, respectively.  
\begin{figure*}[htbp]
\centering
\includegraphics[width=1\textwidth]{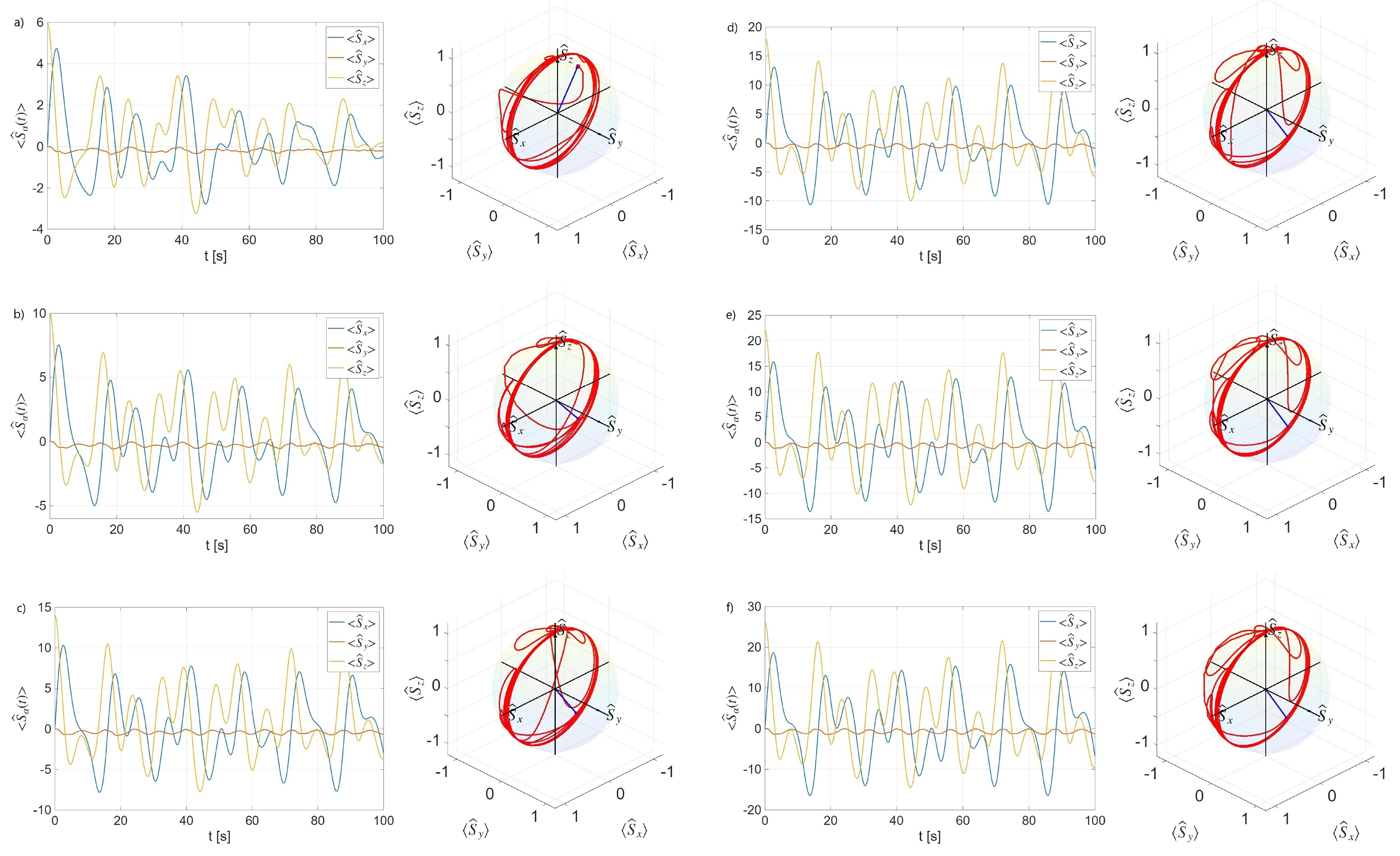}
\caption{\label{fig:wide} Rotating frame evolution of the spin expectation values over 100 s for a two-spin system and the corresponding total magnetization on the Bloch sphere after 100 s, for different Floquet spaces, when $B_{0}=1$, $B_{1}=0.5$, $J=1$, and $D=1$. The Fourier space range corresponds to different truncation orders, having the Floquet index m = 1 (a), 2 (b), 3 (c), 4 (d), 5 (e), and 6 (f).}
\label{fig:2}
\end{figure*}

Overall, these results establish that a moderate Floquet truncation is sufficient to obtain accurate and stable dynamics, while also highlighting the necessity of a convergence-controlled Floquet treatment. This validates the use of an operator-based Floquet-space approach for analyzing driven interacting spin systems and provides a reliable foundation for extending the analysis to more complex cases.

\subsection{Full Floquet theory and approximate Floquet-engineering approaches - a comparison}

Figure~3 compares the rotating-frame dynamics obtained from the first-order Floquet--Magnus approximation with those calculated using the full Floquet formalism for a two-spin system with open boundary conditions, when \(B_0=J=D=1\), and \(B_1=0.2\), \(0.5\), and \(1\). 

Within the approximate treatment [Figs.~3(a)--3(c)], the magnetization components exhibit smooth and regular oscillations. Increasing \(B_1\) mainly increases their frequency, while the qualitative character of the motion remains nearly unchanged. Accordingly, the Bloch-sphere representations consist of simple and closed trajectories, suggesting an almost single-frequency collective rotation. The full Floquet calculations [Figs.~3(d)--3(f)] reveal substantially richer dynamics. The expectation values display amplitude modulations, beating patterns, and several characteristic frequencies, which generate complex and intersecting trajectories on the Bloch sphere. For \(B_1=0.2\), the magnetization remains predominantly localized near the upper hemisphere, although additional loops and irregularities are already present. At \(B_1=0.5\), the trajectory explores a larger region of the sphere, while for \(B_1=1\), where \(B_0\), \(B_1\), \(J\), and \(D\) have equal magnitudes, the motion becomes strongly multifrequency and extends over both hemispheres. Therefore, increasing \(B_1\) does not simply accelerate the spin evolution but also modifies the interference among the contributing Floquet modes.

The discrepancy observed in Fig.~3 should be interpreted as a combination of boundary-sensitive effects and a consequence of comparable energy scales. For the two-spin system, which necessarily represents an open finite chain in the present model, the absence of periodic closure enhances the influence of the interactions and of the higher-order processes neglected by the first-order Floquet--Magnus approximation. The same qualitative disagreement is found for the three-spin chain with open boundary conditions (Appendix A, Figure A3). By contrast, for three spins with periodic boundary conditions, the approximate and full Floquet results become considerably more similar Appendix A, Figure A4), indicating that periodic closure partially restores regular collective dynamics and reduces the impact of the neglected contributions. These results demonstrate that the reliability of the approximate Floquet description depends strongly not only on the energies magnitude, but also on the boundary conditions, with the full Floquet treatment becoming particularly important for accurately describing finite open spin systems or systems where the driving is higher than the other interactions.

\begin{figure*}[htbp]
\centering
\includegraphics[width=1\textwidth]{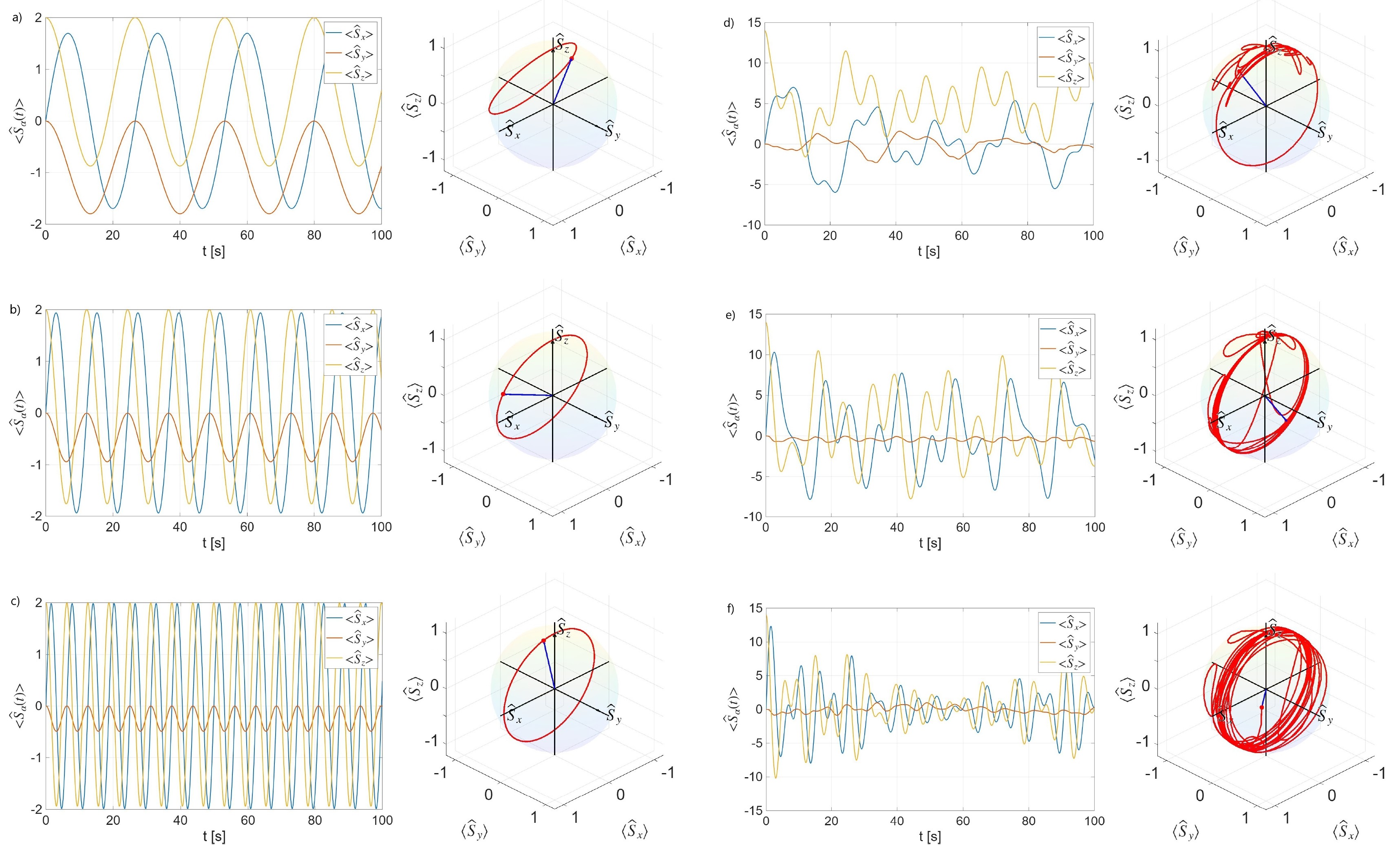}
\caption{\label{fig:wide} Rotating frame evolution of the expectation values over 100 s for a two-spin
system and the corresponding total magnetization on the Bloch sphere after 100 s, obtained using Floquet engineering (a-c), and full Floquet theory (d-f), when $B_{0}=J=D=1$, and $B_{1}=0.2$ (a,d), 0.5 (b,e), and 1 (c,f)} 
\label{fig:3}
\end{figure*}

\subsection{Non-interacting limit}

As a first physical benchmark, we consider the non-interacting limit in which both the isotropic exchange coupling and the Dzyaloshinskii–Moriya interaction are set to zero. In this regime, the driven dynamics is governed solely by the external static and transverse fields, and the system reduces to independent spins undergoing coherent evolution under periodic driving. This limit provides a well-understood reference case, allowing the validity of the Floquet implementation to be verified against expected driven-spin behavior and serving as a baseline for identifying interaction-induced effects. By establishing the characteristic oscillation frequencies and magnetization trajectories in the absence of spin–spin couplings, this section sets the stage for a clear comparison with the interacting regimes discussed below.

\begin{figure*}[htbp]
\centering
\includegraphics[width=1\textwidth]{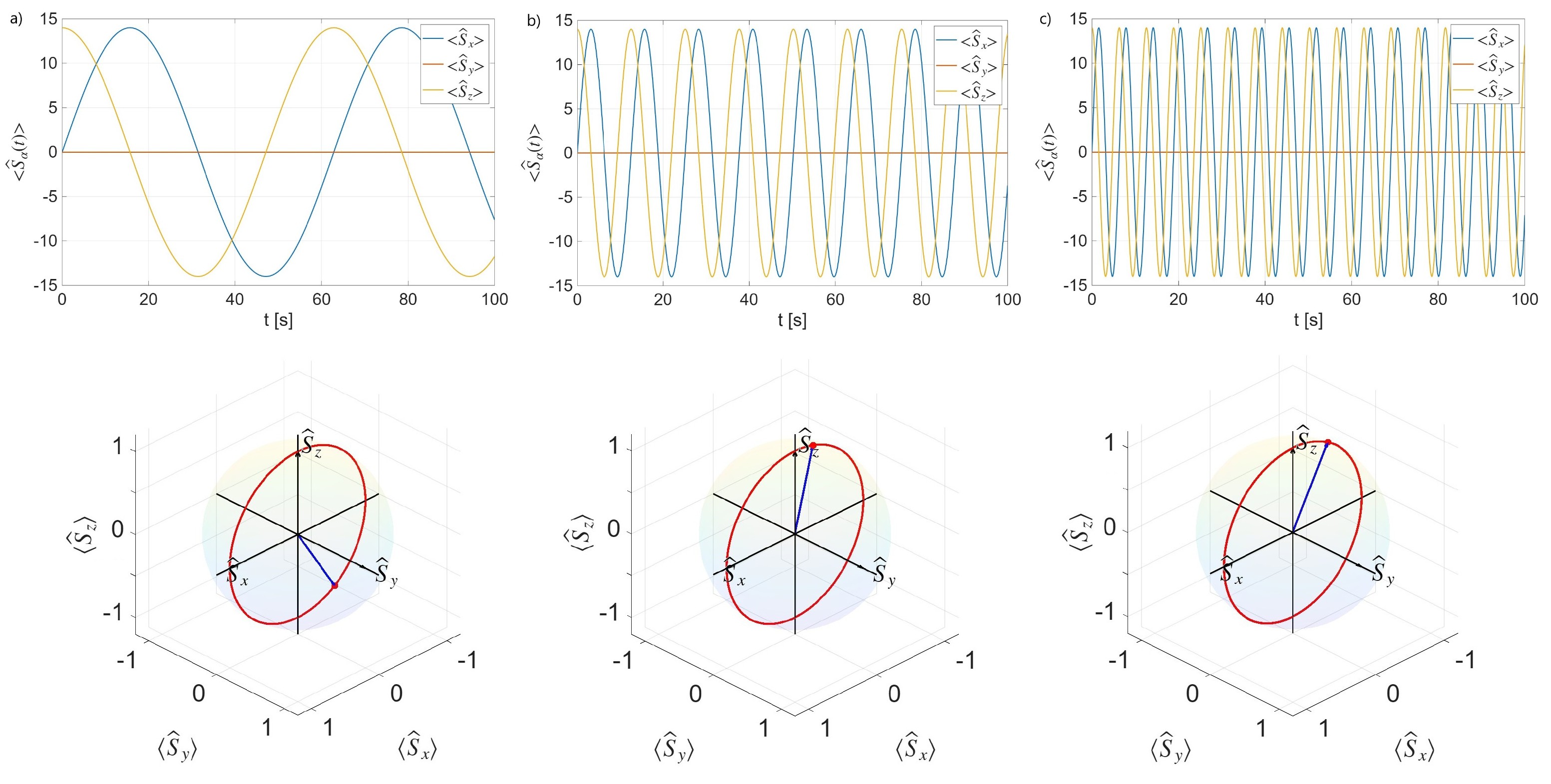}
\caption{\label{fig:wide} Rotating frame evolution of the spin expectation values over 100 s for a two-spin system for three driving amplitudes in frequency units $[B_{0}, B_{1}]=[1, 0.1]$ (a), $[B_{0}, B_{1}]=[1, 0.5]$ (b), and $[B_{0}, B_{1}]=[1, 1]$ (c), and the corresponding total magnetization represented on the Bloch sphere after 100 s}
\label{fig:4}
\end{figure*}

Figure 4 presents the evolution of the spin expectation values over 100 s and the corresponding total magnetization trajectories on the Bloch sphere for a two-spin system in the non-interacting limit, for three driving amplitudes $[B_{0}, B_{1}]=[1, 0.1]$, $[1, 0.5]$, and $[1, 1]$. In this regime, the spins evolve independently under the combined action of the static and transverse fields, and the dynamics is governed solely by the external driving.

For the weakest oscillating field, $[B_{0}, B_{1}]=[1, 0.1]$ (panel a), the spin components exhibit slow, regular oscillations with a well-defined phase relation between $\langle\hat{S}_{x}\rangle$ and $\langle\hat{S}_{z}\rangle$, while $\langle\hat{S}_{y}\rangle=0$  throughout the evolution. The corresponding Bloch-sphere trajectory is a smooth circular orbit, characteristic of coherent precession in an effective field dominated by the static component $B_{0}$. Increasing the field strength to $B_{1}=0.5$ (panel b) leads to a clear increase in the oscillation frequency of the spin expectation values, while preserving their overall coherence and symmetry. For the strongest magnetic field considered, $B_{1}=1$ (panel c), the oscillations become significantly faster, yet remain strictly periodic and stable over the entire time window. The Bloch-sphere representation shows a well-defined circular orbit with increased curvature, indicating rapid coherent rotations driven by the transverse field. Importantly, despite the substantial change in oscillation frequency, the qualitative structure of the trajectory remains unchanged, confirming the absence of interaction-induced distortions.

These simulations were also performed for the three-spin systems and presented in Appendix A (Figures A5 and A6). Comparing the results obtained for the non-interacting three-spin system with those obtained for the two-spin system, one can observe that the driven dynamics are qualitatively identical in all cases. For all driving amplitudes considered, the spin expectation values display purely coherent, single-frequency oscillations, with the oscillation periodicity determined by the transverse driving field $B_{1}$. The relative phase relations between $\langle\hat{S}_{x}\rangle$ and $\langle\hat{S}_{z}\rangle$, as well as $\langle\hat{S}_{y}\rangle=0$, are preserved when moving from two to three spins. This equivalence is further confirmed by the Bloch-sphere representations, which show the same closed circular trajectories for both systems. Increasing the number of spins by one does not introduce additional frequency components, distortions, or damping-like features in the absence of spin–spin interactions. Consequently, the collective magnetization behaves as a simple sum of independent spin contributions, and the Floquet-engineered response remains governed solely by the external driving fields. These observations demonstrate that, in the non-interacting limit, the extension from two to three spins does not modify the qualitative structure of the driven dynamics.

Overall, it was demonstrated that in the non-interacting limit, the driven dynamics is fully coherent and characterized by single-frequency behavior, with the transverse field $B_{1}$ controlling the oscillation periodicity and the geometry of the Bloch-sphere trajectory. This well-understood behavior provides a robust reference point for identifying the qualitative and quantitative modifications introduced by exchange and chiral interactions in the interacting regimes discussed below.

\subsection{Spin system subjected to symmetric direct exchange coupling}
Having established the baseline behavior in the non-interacting limit, the effect of isotropic exchange coupling on the coherently driven spin dynamics is further investigated by setting the Dzyaloshinskii–Moriya interaction to zero and comparing the extreme cases $J=0$ and $J=1$. For this purpose, the simulations are performed for fixed driving parameters $[B_{0}, B_{1}]=[1, 0.5]$, for a two-spin system, allowing the influence of exchange coupling to be isolated from changes in the external driving (Figure 5).

\begin{figure*}[htbp]
\centering
\includegraphics[width=0.8\textwidth]{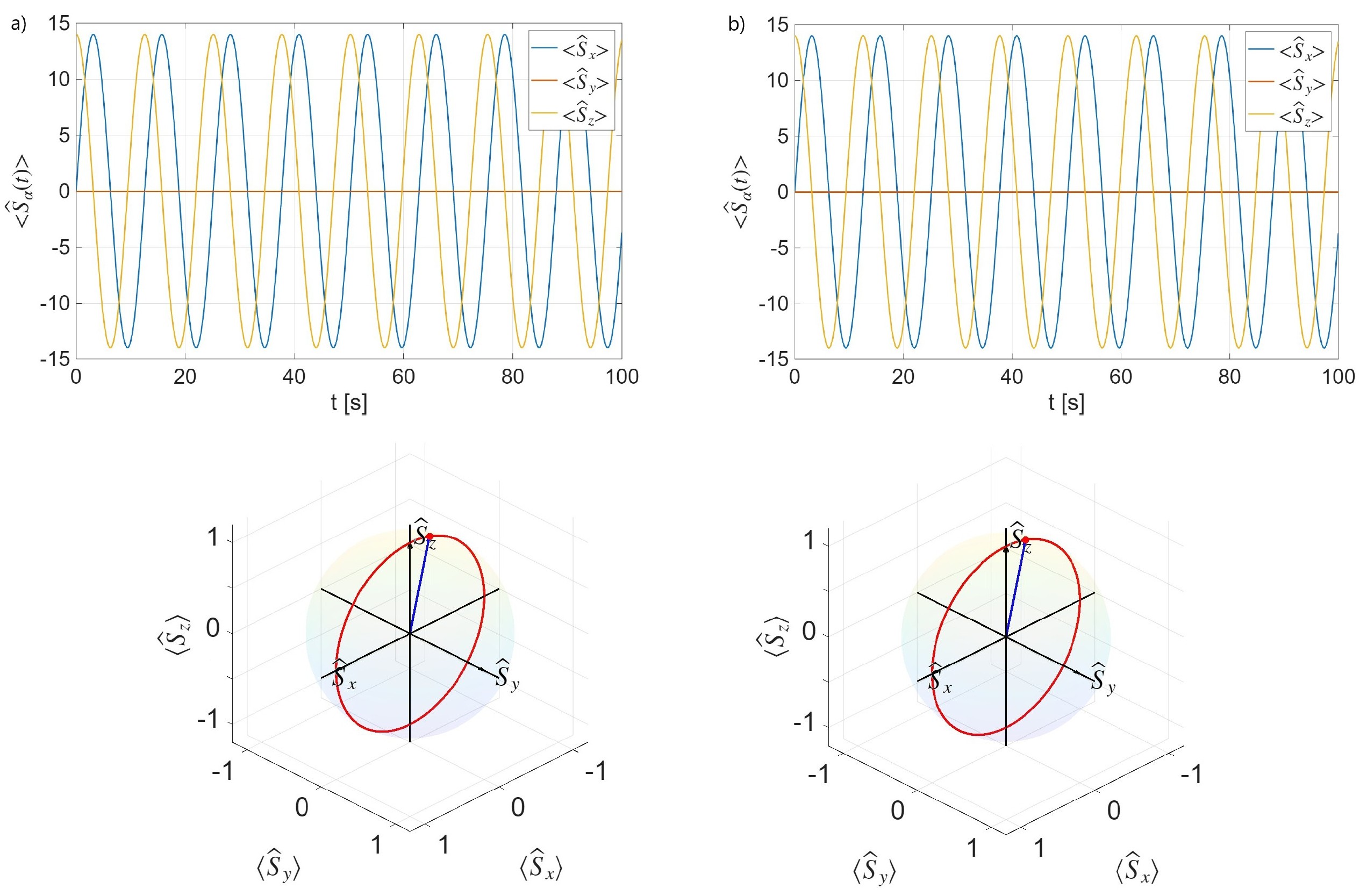}
\caption{\label{fig:wide} Rotating frame evolution of the spin expectation values over 100 s for a two-spin system, when $[B_{0}, B_{1}]=[1, 0.5]$, DMI interaction is neglected, the exchange coupling is set to 0 (a) or 1 (b), and the corresponding total magnetization represented on the Bloch sphere after 100 s}
\label{fig:5}
\end{figure*}

A direct comparison of the two cases reveals no observable difference in the driven dynamics when the exchange interaction is switched on. The time evolution of $\langle\hat{S}_{x}(t)\rangle$, and $\langle\hat{S}_{z}(t)\rangle$ exhibits identical oscillation amplitudes, frequencies, and phase relations for $J=0$ and $J=1$, and $\langle\hat{S}_{y}(t)\rangle=0$ throughout the evolution. Likewise, the Bloch-sphere representations show the same closed, circular trajectories, indicating that the collective magnetization follows an identical coherent path in both cases. This behavior can be understood from symmetry considerations. The isotropic exchange interaction conserves the total spin and commutes with the collective spin operators under the chosen initial polarization (i.e. $[\hat{H}_{J}, \hat{S}_{x}]=[\hat{H}_{J}, \hat{S}_{y}]=[\hat{H}_{J}, \hat{S}_{z}]=0$, where $\hat{H}_{J}=J\sum_{i,j}(\hat{S}_{ix}\hat{S}_{jx}+\hat{S}_{iy}\hat{S}_{jy}+\hat{S}_{iz}\hat{S}_{jz})$). Consequently, although the exchange term modifies the internal structure of the two-spin eigenstates, it does not contribute to the time evolution of the collective spin expectation values probed here. As a result, the driven dynamics remains governed solely by the external fields, and the system behaves effectively as in the non-interacting case with respect to these observables. This result remains the same, whatever the number of spins is. Therefore, isotropic exchange coupling alone is insufficient to alter the coherent Floquet-driven magnetization dynamics under the present conditions. This finding highlights the necessity of symmetry-breaking interactions, such as the Dzyaloshinskii–Moriya interaction, to generate qualitatively new dynamical features beyond simple coherent rotations, which are examined in the following sections.

\subsection{The effect of the Dzyaloshinskii–Moriya interaction in the absence of symmetric direct exchange coupling}

The role of the Dzyaloshinskii–Moriya interaction is further isolated by setting the isotropic exchange coupling to zero and examining the driven dynamics as a function of the DMI strength. Unlike exchange interaction, which preserves rotational symmetry and does not affect the collective magnetization dynamics under the chosen conditions, the DMI introduces an antisymmetric spin–spin coupling that breaks inversion symmetry and mixes spin components. As a result, it is expected to generate qualitatively new dynamical features under periodic driving. By focusing on the DMI-only regime, this subsection provides a minimal setting to investigate how chiral interactions reshape coherent spin motion in the absence of competing symmetric couplings. This analysis allows the direct identification of DMI-induced effects—such as the emergence of additional transverse magnetization components, deformation of Bloch-sphere trajectories, and the onset of multi-frequency dynamics—and establishes a reference point for understanding the more complex behavior arising from the combined action of exchange and DMI interactions discussed in the following section.

Figures 6 and 7 present the driven dynamics of a three-spin system in the absence of isotropic exchange coupling, for increasing values of the Dzyaloshinskii–Moriya interaction (DMI), under open boundary conditions (Fig. 6) and periodic boundary conditions (Fig. 7), respectively. All simulations are performed for fixed driving parameters $[B_{0}, B_{1}]=[1, 0.5]$. For completeness, the corresponding results for the two-spin system are provided in Appendix A (Figure $A7$), as they are found to be qualitatively identical to those obtained for the three-spin system with open boundary conditions.

For vanishing DMI (panels a), the dynamics reduces to the non-interacting case discussed previously: the spin expectation values exhibit purely coherent oscillations governed by the driving field, and the Bloch-sphere trajectories correspond to closed, tilted elliptical orbits. As the DMI strength is increased (panels b–f), systematic and robust modifications of the dynamics emerge. A finite transverse $\langle\hat{S}_{y}\rangle$ component develops, while the amplitudes of $\langle\hat{S}_{x}\rangle$ and $\langle\hat{S}_{z}\rangle$ are progressively reduced. In the time domain, the oscillations become increasingly modulated, indicating the onset of multi-frequency dynamics induced by the chiral interaction. These effects are reflected in the Bloch-sphere representations. For small DMI, the magnetization follows a slightly distorted but still well-defined elliptical trajectory. With increasing DMI, the trajectory thickens and gradually evolves into a more complex, multi-loop structure, indicating that the magnetization no longer executes simple coherent rotations. Instead, the dynamics explores a broader region of the Bloch sphere as polarization is continuously transferred into correlated multi-spin channels.

\begin{figure*}[htbp]
\centering
\includegraphics[width=1\textwidth]{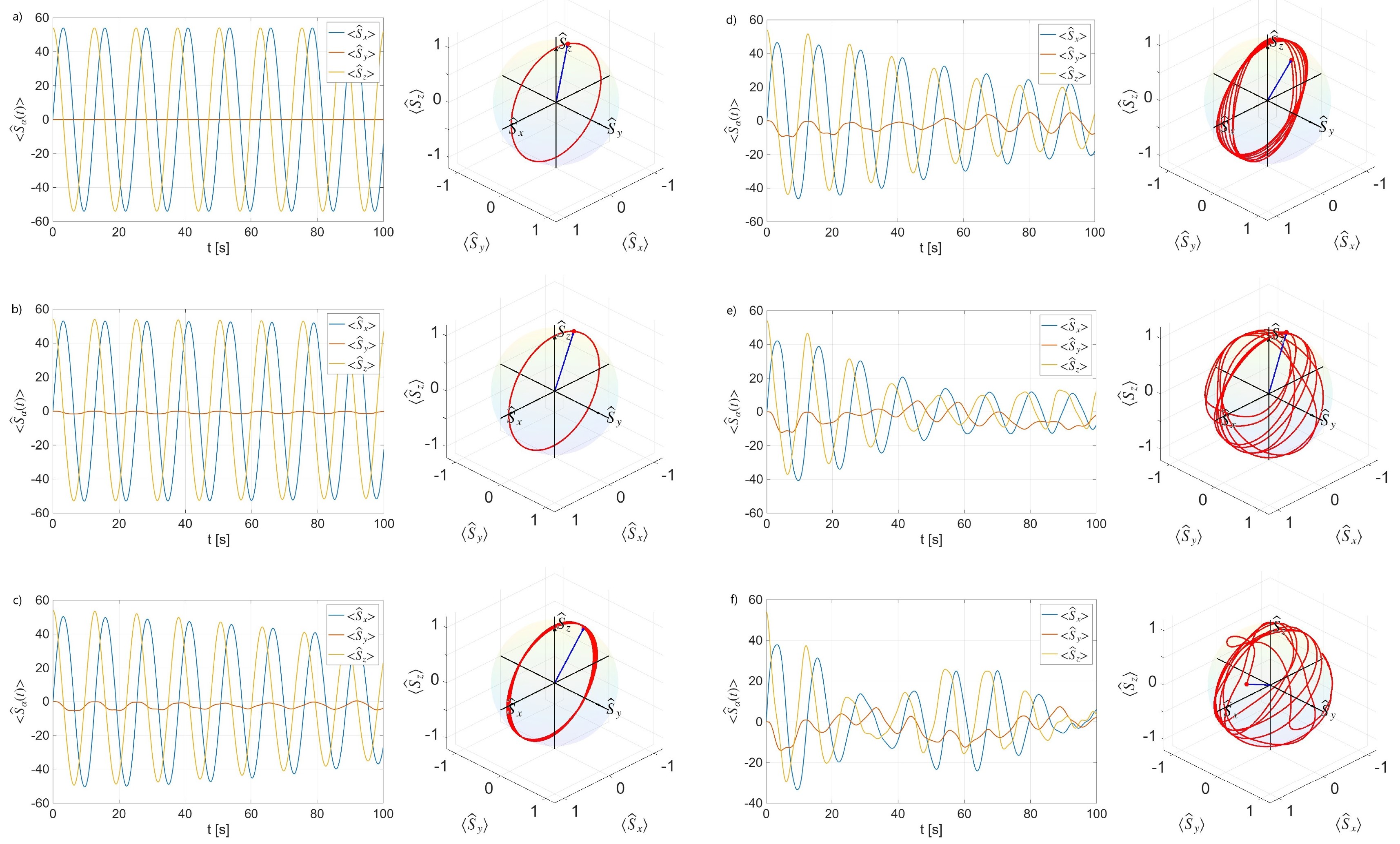}
\caption{\label{fig:wide} Rotating frame evolution of the spin expectation values over 100 s for a three-spin system with open boundary condition, when $[B_{0}, B_{1}]=[1, 0.5]$, the exchange interaction is neglected, the DMI is set to 0 (a), 0.2 (b), 0.4 (c), 0.6 (d), 0.8 (e), and 1 (f), and the corresponding total magnetization represented on the Bloch sphere after 100 s}
\label{fig:6}
\end{figure*}

\begin{figure*}[htbp]
\centering
\includegraphics[width=1\textwidth]{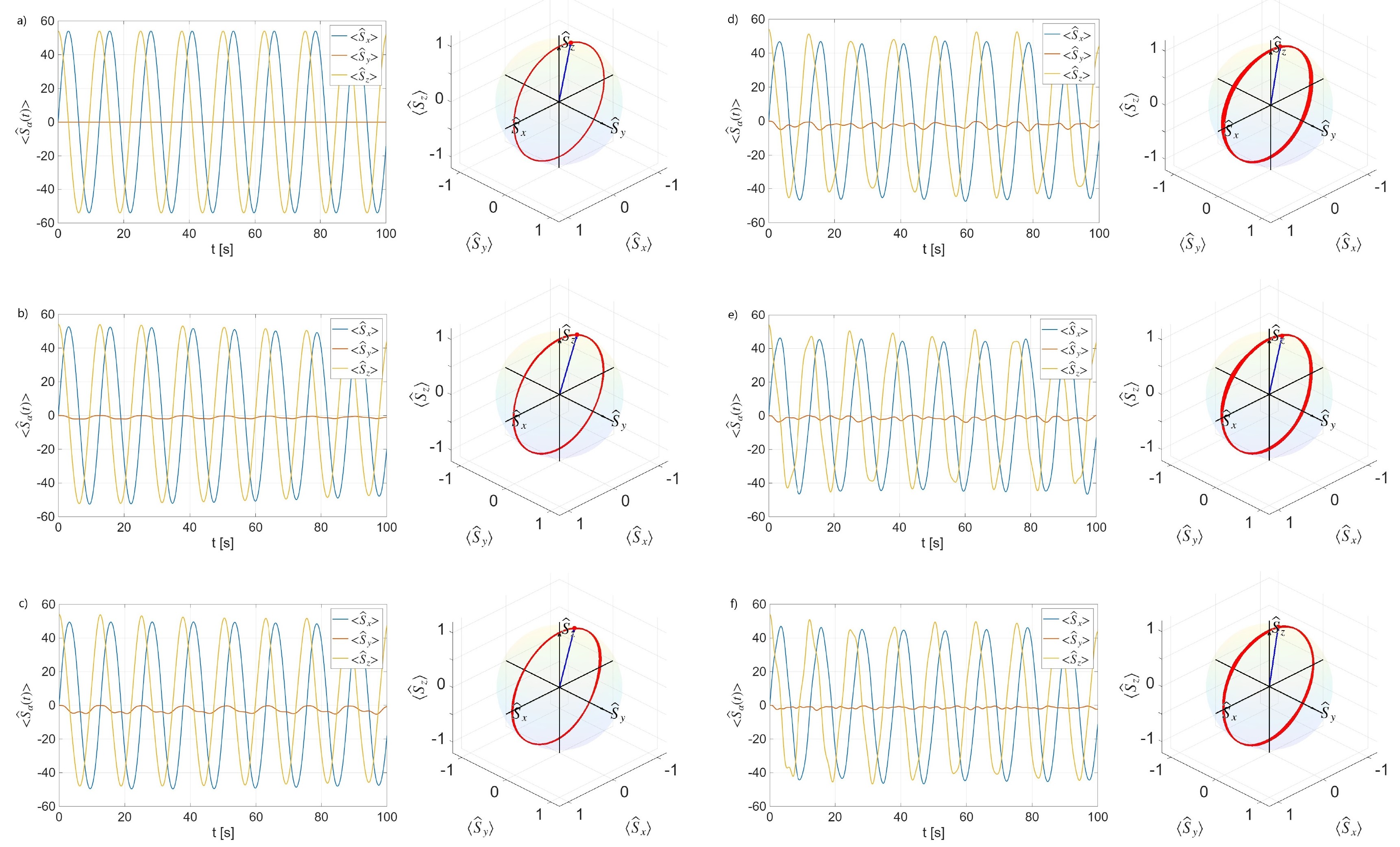}
\caption{\label{fig:wide} Rotating frame evolution of the spin expectation values over 100 s for a three-spin system with periodic boundary condition, when $[B_{0}, B_{1}]=[1, 0.5]$, the exchange interaction is neglected, the DMI is set to 0 (a), 0.2 (b), 0.4 (c), 0.6 (d), 0.8 (e), and 1 (f), and the corresponding total magnetization represented on the Bloch sphere after 100 s}
\label{fig:7}
\end{figure*}

A clear distinction emerges when comparing open and periodic boundary conditions for the three-spin system. Under open boundary conditions (Fig. 6), the DMI-induced distortions are more pronounced, and the Bloch-sphere trajectories exhibit stronger thickening and loss of regularity as the DMI strength increases. In contrast, for periodic boundary conditions (Fig. 7), the magnetization trajectories remain more regular and closer to closed orbits, even at larger DMI values. The time-domain signals show reduced modulation compared to the open-chain case, indicating a partial suppression of boundary-induced correlation effects. This difference can be understood in terms of interaction topology. In the open chain, the lack of translational symmetry and the presence of edge spins enhance the sensitivity to chiral interactions, facilitating the redistribution of magnetization into non-uniform and correlated modes. In the periodic three-spin ring, all spins are equivalent and experience the same local environment, which partially stabilizes the collective motion and mitigates the impact of DMI on the global magnetization dynamics. Therefore, in the absence of exchange coupling, the DMI alone is sufficient to qualitatively reshape the driven dynamics, generating multi-frequency behavior and breaking the simple Bloch-sphere rotation picture. While the minimal two-spin system already captures the essential DMI-induced effects, extending the system to three spins reveals the additional role of boundary conditions, highlighting the importance of interaction topology in Floquet-engineered chiral spin dynamics.

\subsection{Realistic picture: combine effect of the exchange coupling and Dzyaloshinskii–Moriya interaction}

\begin{figure*}[htbp]
\centering
\includegraphics[width=1\textwidth]{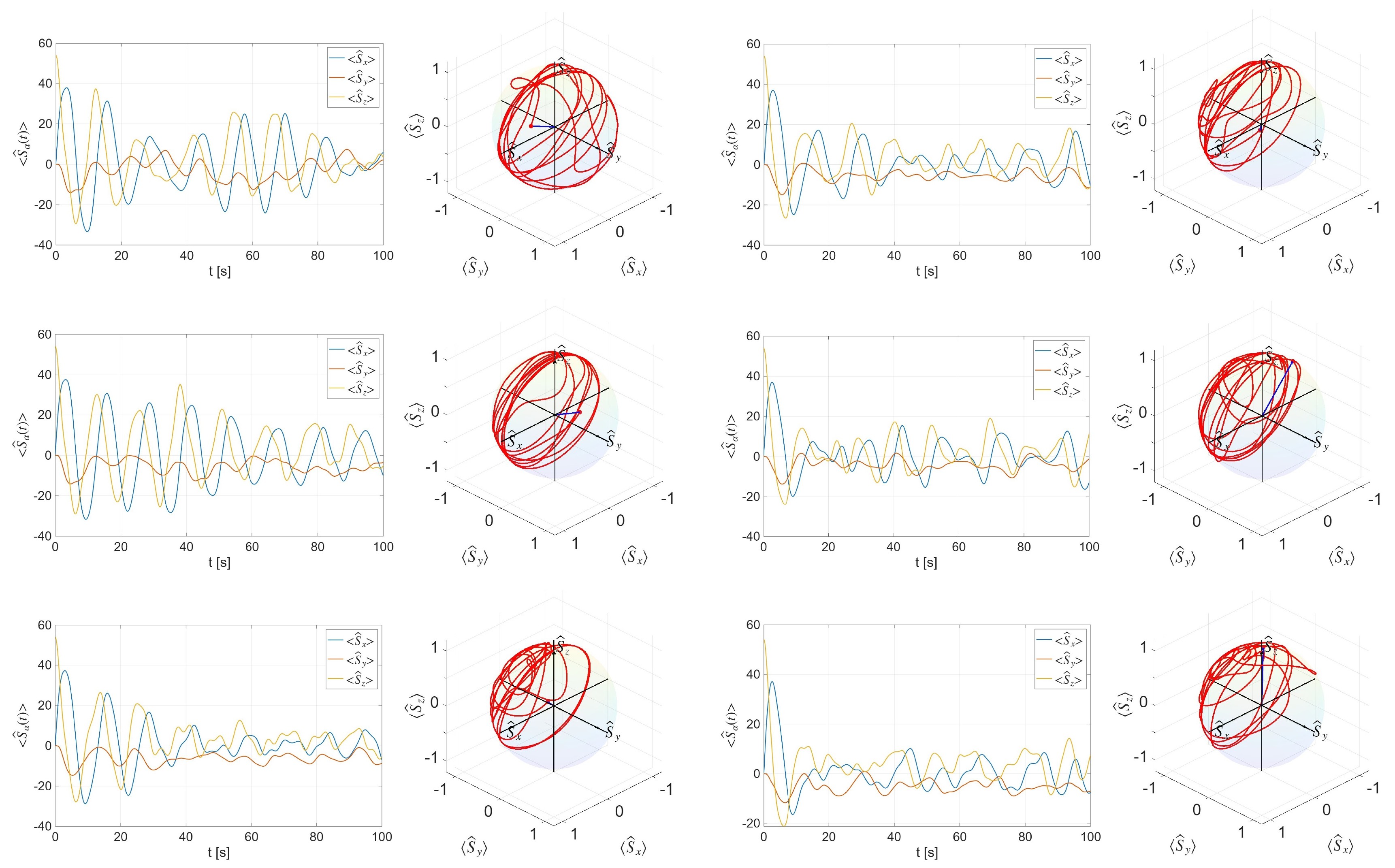}

\caption{\label{fig:wide} Rotating frame evolution of the spin expectation values over 100 s for a three-spin system with open boundary condition, when $[B_{0}, B_{1}]=[1, 0.5]$, the DMI is set to 1, the exchange interaction is 0 (a), 0.2 (b), 0.4 (c), 0.6 (d), 0.8 (e), and 1 (f), and the corresponding total magnetization represented on the Bloch sphere after 100 s}
\label{fig:8}
\end{figure*}

The most realistic scenario is when isotropic exchange coupling and the DMI interaction act simultaneously in a coherently driven spin system. While the previous sections isolated the individual effects of exchange and chiral interactions, their coexistence is expected to generate nontrivial dynamics due to the competition between symmetric and antisymmetric couplings. In this regime, exchange tends to promote collective spin alignment, whereas the DMI introduces chirality and mixes spin components, leading to complex multi-frequency behavior under periodic driving. By analyzing this combined interaction regime, we aim to elucidate how realistic spin–spin interactions reshape coherent Floquet-driven dynamics beyond simple rotation-based control schemes.

Figures 8 and 9 show the evolution of the spin expectation values and the corresponding total magnetization trajectories on the Bloch sphere for a three-spin system subjected simultaneously to Dzyaloshinskii–Moriya interaction ($DMI = 1$) and isotropic exchange coupling, under open boundary conditions (Fig. 8) and periodic boundary conditions (Fig. 9), respectively. The driving parameters, in frequency units, are fixed to $[B_{0}, B_{1}]=[1, 0.5]$, while the exchange coupling is varied from $J=0$ to $J=1$. Similar simulations were performed for the two-spin system (Figure $A8$,  Appendix A), the results being qualitatively similar to those obtained for the three-spin system with open boundary conditions.

For open boundary conditions (Fig. 8), the combined presence of DMI and exchange leads to strongly perturbed dynamics. In the absence of exchange ($J=0$, panel a), the dynamics is already multi-frequency and irregular, reflecting the dominant role of the chiral interaction discussed in the previous section. As the exchange coupling increases (panels b–f), our calculations show a counterintuitive result: the time-domain spin expectation values exhibit progressively stronger amplitude modulation and partial suppression of coherent oscillations. Correspondingly, the Bloch-sphere trajectories evolve from distorted elliptical paths into increasingly dense, multi-loop structures that fill a significant portion of the Bloch sphere. This behavior indicates enhanced redistribution of magnetization into correlated multi-spin channels and a breakdown of simple coherent rotation due to the competition between symmetric (exchange) and antisymmetric (DMI) interactions in an open geometry.  In contrast, the behavior under periodic boundary conditions (Fig. 9) is markedly different. For all values of the exchange coupling, the spin expectation values remain regular and nearly single-frequency, and the Bloch-sphere trajectories maintain well-defined, closed, elliptical orbits. Interestingly, increasing $J$ does not introduce visible distortions or additional complexity in the magnetization dynamics. Instead, the periodic three-spin ring appears to stabilize the collective motion, effectively suppressing the DMI-induced multi-frequency behavior observed in the open chain.

\begin{figure*}[htbp]
\centering
\includegraphics[width=1\textwidth]{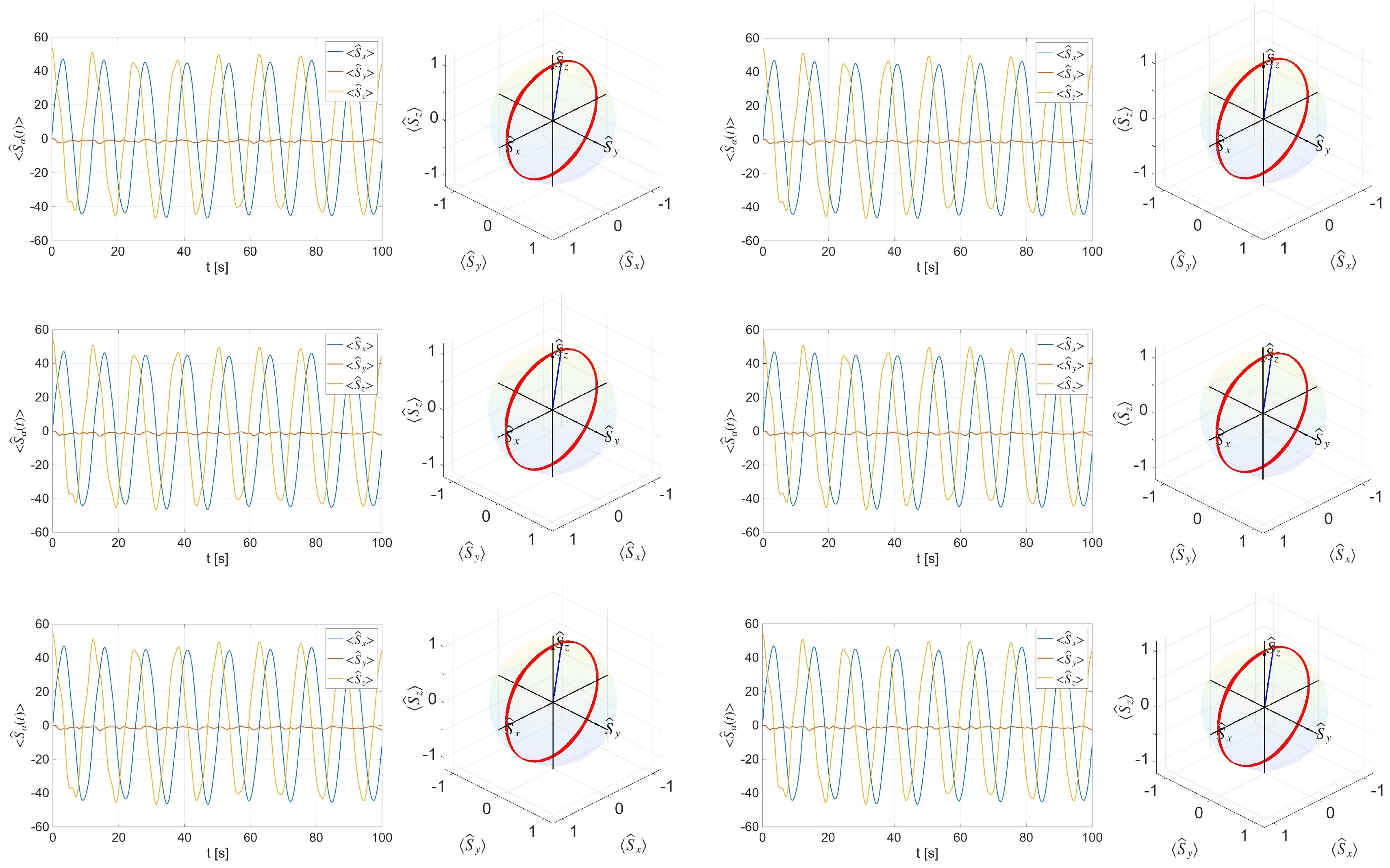}

\caption{\label{fig:wide} Rotating frame evolution of the spin expectation values over 100 s for a three-spin system with periodic boundary condition, when $[B_{0}, B_{1}]=[1, 0.5]$, the DMI is set to 1, the exchange interaction is 0 (a), 0.2 (b), 0.4 (c), 0.6 (d), 0.8 (e), and 1 (f), and the corresponding total magnetization represented on the Bloch sphere after 100 s}
\label{fig:9}
\end{figure*}

This pronounced difference highlights the crucial role of interaction topology. In the open chain, broken translational symmetry and the presence of edge spins enhance the sensitivity to competing interactions, allowing DMI and exchange to generate complex, strongly correlated dynamics jointly. In the periodic ring, all spins are equivalent and experience a symmetric interaction environment, which constrains the redistribution of magnetization and preserves coherent collective motion even when both interactions are present. The combined effect of exchange and DMI is highly boundary-dependent. While the coexistence of these interactions leads to strongly perturbed, non-Bloch-like dynamics in open spin chains, periodic boundary conditions largely restore regular coherent behavior. This finding underscores that realistic modeling of driven chiral spin systems must account not only for interaction strengths but also for system topology, as boundary effects can qualitatively reshape Floquet-engineered spin dynamics. These effects are particularly pronounced and clearly distinguishable in the extremely low-dimensional regime that we addressed here, whereas in larger systems they may be partially obscured by additional sources of complex quantum fluctuations such as frustration or competing exchange mechanisms leading to non-collinear spin or spin-spin correlation features. For example, we explicitly show here that, in short chains, periodic boundary conditions enhance spin coherence during the driven dynamics. In contrast, for larger two-dimensional skyrmionic lattices, it has been recently shown \cite{Tiusan2025} that the time coherence under periodic boundary conditions is reduced compared to the corresponding system with open boundary conditions, due to the lack of topological stabilization in the translationally invariant setting enforced by PBC. \textit{Appendix B} discusses the physical implications of the predicted Floquet trajectories and DMI-induced distortions of experimentally accessible observables, together with prospects for implementing Floquet-controlled quantum operations in low-dimensional magnetic and skyrmionic systems.

\subsection{General overview and outlook}

\begin{figure*}
 \centering
        \includegraphics[width=0.8\textwidth]{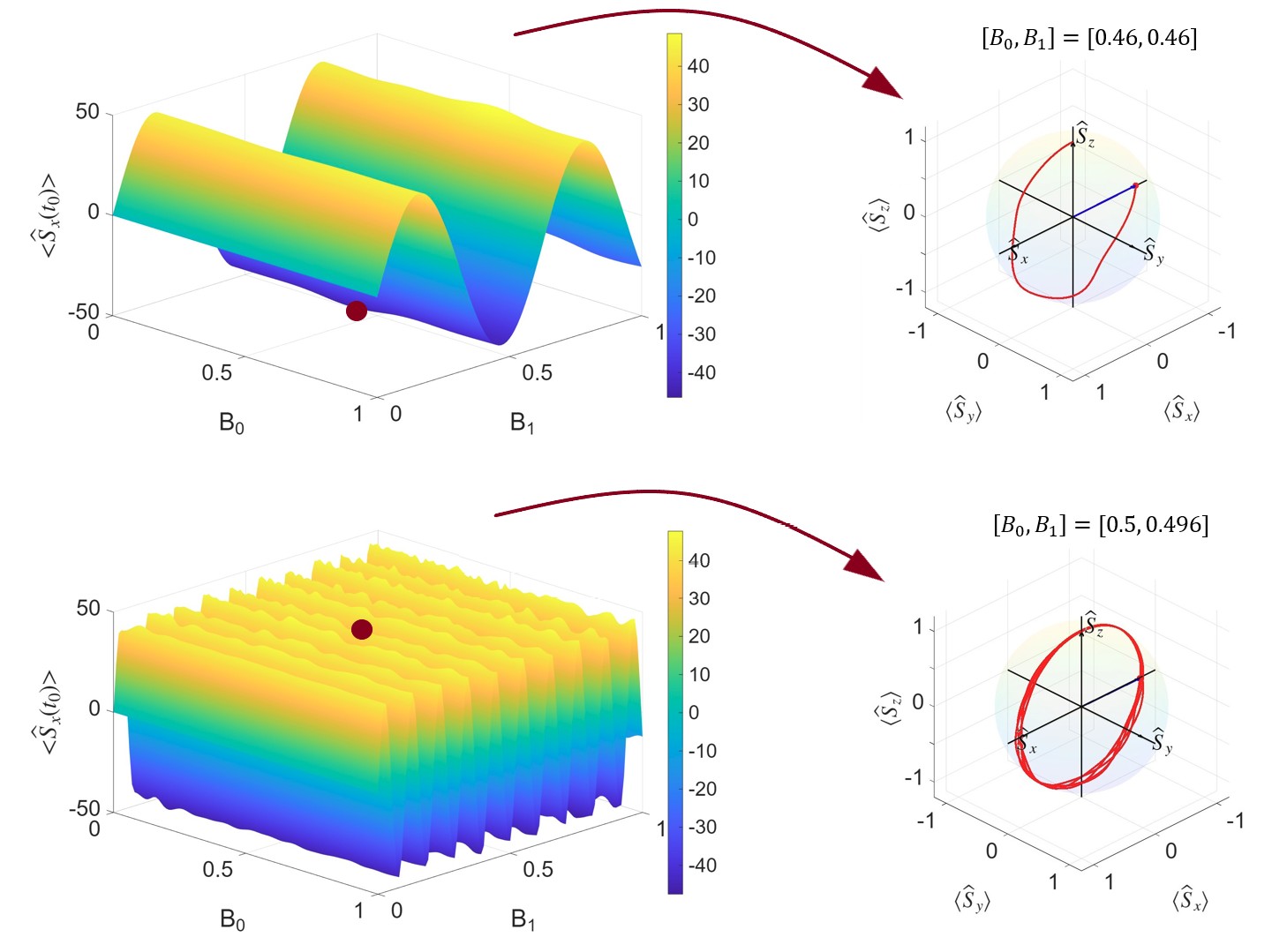}
\caption{\label{fig:wide}Direct correlation between the Floquet control landscape and the resulting 
spin dynamics after 10 s (a) and 60 s (b) of evolution. Surface maps of the transverse spin component $\langle \hat{S}_x(t_{0}) \rangle$ 
are shown together with the corresponding Bloch-sphere trajectories of the total magnetization 
for parameter combinations $(B_0, B_1)$ selected from the control surfaces, as indicated in the figure. These points 
indicate driving conditions that produce large transverse magnetization, while the adjacent 
Bloch trajectories confirm the associated longitudinal-to-transverse spin rotation. This combined representation illustrates how the global parameter-space structure predicts the qualitative nature of the Floquet-driven dynamics in the interacting chiral spin system.}
\label{fig:10}
\end{figure*}

This section presents a global control-oriented perspective. For this purpose, the combined effects of coherent driving and interactions are explored to achieve targeted spin rotations with maximum coherence degree, in direct analogy with pulse design strategies used in nuclear magnetic resonance (NMR). Surface maps of the longitudinal spin component $\langle\hat{S}_{z}(t_{0})\rangle$ are evaluated at two different evolution times, $t=10$ s, and 60 s, as a function of the static and transverse field amplitudes $B_{0}$ and $B_{1}$, varied continuously over the interval $[0, 1]$. The analysis is performed for a three-spin system with periodic boundary conditions, as a general example, in a realistic interaction regime where the Dzyaloshinskii–Moriya interaction is finite and competes with isotropic exchange coupling ($DMI=1$, $J=0.5$) (Figure 10, left panel).

Since the initial magnetization is oriented along the z-axis, the appearance of large positive or negative values of $\langle\hat{S}_{x}(t_{0})\rangle$ directly signals an effective $\pi/2$ rotation of the spins from the longitudinal plane into the transverse direction, in close analogy with a $\pi/2$ pulse in NMR. In contrast, regions where $\langle\hat{S}_{x}(t_{0})\rangle$ remains close to zero correspond to parameter combinations for which the magnetization stays predominantly in the longitudinal plane. At the shorter evolution time ($t_{0}$=10 s, Figure 10 a), the surface map exhibits broad, smooth regions of positive and negative $<\hat{S}_{x}(t_{0})>$, indicating that NMR-like $\pi/2$ rotations can be achieved over extended intervals of $B_{1}$ for moderate values of $B_{0}$. This suggests a degree of robustness with respect to parameter variations, similar to broadband excitation conditions in conventional NMR pulse design. At the longer evolution time ($t_{0}=60$ s, Figure 10 b), the control landscape becomes more structured, displaying finer oscillatory features as a function of both $B_{0}$ and $B_{1}$. These modulations reflect the accumulation of multi-frequency Floquet phases induced by the competing exchange and chiral interactions. Nevertheless, well-defined ridges of large positive and negative $<\hat{S}_{x}(t_{0})>$ persist, demonstrating that targeted $\pi/2$-like rotations remain accessible, albeit with increased sensitivity to the precise choice of driving parameters.

Two examples of a $\pi/2$ rotation of the spins from the z-axis onto the negative part of the x-axis are obtained when $B_{0}=B_{1}=0.46$ after 10 s, and for $[B_{0}, B_{1}]=[0.5, 0.496]$ after 60 s, their adjacent Bloch-sphere panels displaying the resulting evolution of the total magnetization (Figure 9, right panel). The corresponding Bloch spheres for these specific values of $[B_{0}, B_{1}]$ are represented next to the surface maps (Figure 10).

More generally, the examples presented above demonstrate that the Floquet-space formalism developed in this work is not restricted to realizing specific $\pi/2$ rotations. By systematically exploring the driving-parameter space and the evolution time, the same approach can be used to design arbitrary controlled spin rotations, including full inversions, partial rotations, and composite trajectories in the presence of interactions. 
These predictive results represent an important step toward the exploration, design, and fundamental understanding of the behaviour and properties of next-generation spintronic devices, including platforms relevant to quantum-technology applications. As noted, our results are expressed in terms of interaction and field parameters scaled in frequency units. Consequently, the realization of a realistic experimental material and the corresponding quantum spintronic device requires engineering spin systems with appropriate interaction strengths ($J$ and $DMI$) and control parameters $B_0$ and $B_1$. Depending on the chosen material platform and device architecture, the relative magnitudes of $DMI$ and the $J$-coupling can be tuned through the underlying material composition and interaction design. 

Then, the type of surface-map analysis whose framework and methodology were developed and illustrated here provides direct guidance for the spin manipulation, e.g. for implementation of quantum gate operations, selecting combinations of field amplitudes and pulse durations that achieve a desired final spin orientation, closely mirroring pulse calibration procedures in magnetic resonance. We underline the important fact that this strategy remains applicable in strongly interacting and chiral regimes, where simple analytical pulse prescriptions fail. As such, the present framework establishes a versatile route toward Floquet-engineered spin control beyond single-frequency or weak-coupling approximations, with potential extensions to multi-pulse sequences, shaped driving fields, and larger spin networks.

\section{Conclusion}
In this work, we developed a systematically converged, operator-based full Floquet-space framework for the description and coherent control of periodically driven interacting electron-spin systems. Adapted from solid-state NMR methodology, the formalism treats the complete time-periodic Hamiltonian in an extended Hilbert--Floquet space and retains the Floquet modes, quasienergy structure, resonant contributions, and higher-order processes that are absent from a first-order Floquet--Magnus effective Hamiltonian. A direct comparison between the two approaches shows that the approximate treatment can predict artificially regular, nearly single-frequency dynamics, whereas the full Floquet solution reveals amplitude modulations, beating patterns, and complex magnetization trajectories.

The two- and three-spin systems considered here provide minimal models in which the separate and combined effects of isotropic exchange, the Dzyaloshinskii--Moriya interaction (DMI), and boundary conditions can be identified. For the chosen initial state, isotropic exchange alone does not alter the collective-spin expectation values, whereas DMI generates a finite transverse component and transforms simple collective rotations into tilted, distorted, and multifrequency trajectories. Most importantly, the disagreement between the approximate and full Floquet descriptions is strongly dependent on the interaction topology. It is pronounced for the two-spin system and the three-spin chain with open boundary conditions, for which boundary-induced inequivalence enhances the influence of interaction-generated correlations and of the processes neglected by the approximate treatment. By contrast, for three spins with periodic boundary conditions, the two approaches yield considerably more similar dynamics, indicating that periodic closure partially restores regular collective motion even in the presence of competing symmetric and antisymmetric interactions. The validity of a low-order Floquet approximation therefore depends not only on the relative energy scales but also critically on the boundary conditions and connectivity of the spin system.

Beyond characterizing the dynamics, we introduced a control-oriented description based on maps of the longitudinal-spin response over the driving-parameter space. These control landscapes identify field amplitudes and evolution times that produce targeted NMR-like operations, including robust $\pi/2$ rotations from an initially longitudinal magnetization. They demonstrate that coherent-control concepts developed in magnetic resonance can be transferred to interacting chiral spin systems. Within the same framework, other controlled rotations can, in principle, be designed by selecting appropriate driving parameters and evolution times.

Our results establish full Floquet theory as a predictive framework for Hamiltonian engineering beyond perturbative Floquet descriptions in finite interacting spin systems. Although the present analysis focuses on the low-dimensional limit, the formalism is not conceptually restricted to two- and three-spin models, with its extension being limited primarily by computational cost. It can therefore be generalized to larger spin networks, different interaction geometries, higher-dimensional architectures, and tailored driving protocols. Such developments may support the design of coherent-control strategies for quantum information processing, magnonics, and spintronic devices, including quantum operations based on topological spin textures~\cite{Psaroudaki2021,Psaroudaki2023,Xia2023,Petrovic2025}.

\section*{Acknowledgements}
% TODO: include funding information
\paragraph{Funding information}
A.S. acknowledges grant no. 106/06.02.2026 funded by the national fellowship program L'Oreal-UNESCO for Women in Science. C.F. acknowledges financial support from the MCID through the “Nucleu” Program within the National Plan for Research, Development, and Innovation 2022–2027, Project No. PN23 24 01 05.  C. T. acknowledges funding from UEFISCDI via the project “MODESKY” PN-III-P4-ID-PCE-2020-0230-P, grant No.~UEFISCDI: PCE 245/02.11.2021.

\begin{appendix}
\numberwithin{equation}{section}

\section{Additional Full Floquet Theory applications on the two-spin and three-spin systems}

\renewcommand{\thefigure}{A\arabic{figure}}
\setcounter{figure}{0}

This section presents additional full Floquet simulations for the two- and three-spin systems investigated in the main text. These results complement the analysis presented in the manuscript by further illustrating the effects of the driving parameters, spin interactions, and boundary conditions on the collective magnetization dynamics and the corresponding Bloch-sphere trajectories.

\subsection{Floquet-space convergence and truncation order for three-spin system}

Figures $A1$ and $A2$ illustrate the convergence behavior of the Floquet-space formalism for a three-spin system with open boundary conditions (Figure $A1$) and periodic boundary conditions (Figure $A2$). As the truncation order increases, the spin expectation values and Bloch-sphere trajectories progressively stabilize. While low truncation orders produce visible distortions, a Floquet index of \textit{$m \in [-4, 4]$} yield nearly identical dynamics for both cases, confirming numerical convergence and validating the truncation procedure adopted throughout the study.

\begin{figure*}[htbp]
\centering
\includegraphics[width=1\textwidth]{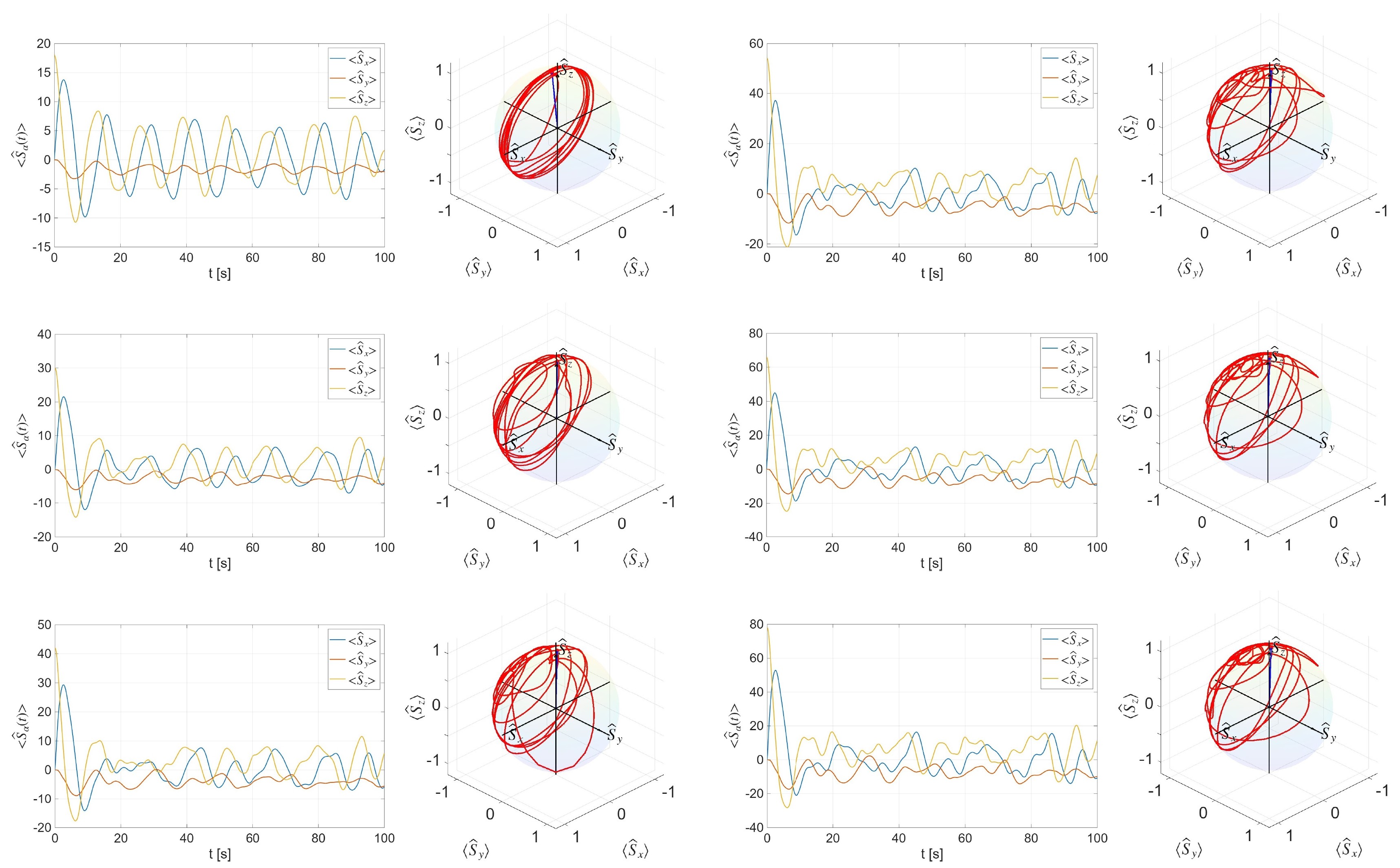}
 \caption{Rotating frame evolution of the spin expectation values over 100 s for a three-spin system with open boundary conditions and the corresponding total magnetization on the Bloch sphere after 100 s, for different Floquet spaces, when $B_{0}=1$, $B_{1}=0.5$, $J=1$, and $DMI=1$. The Fourier space range corresponds to different truncation orders, having the Floquet index m = 1 (a), 2 (b), 3 (c), 4 (d), 5 (e), and 6 (f)}
\label{fig:A1}
\end{figure*}

\begin{figure*}[htbp]
\centering
\includegraphics[width=1\textwidth]{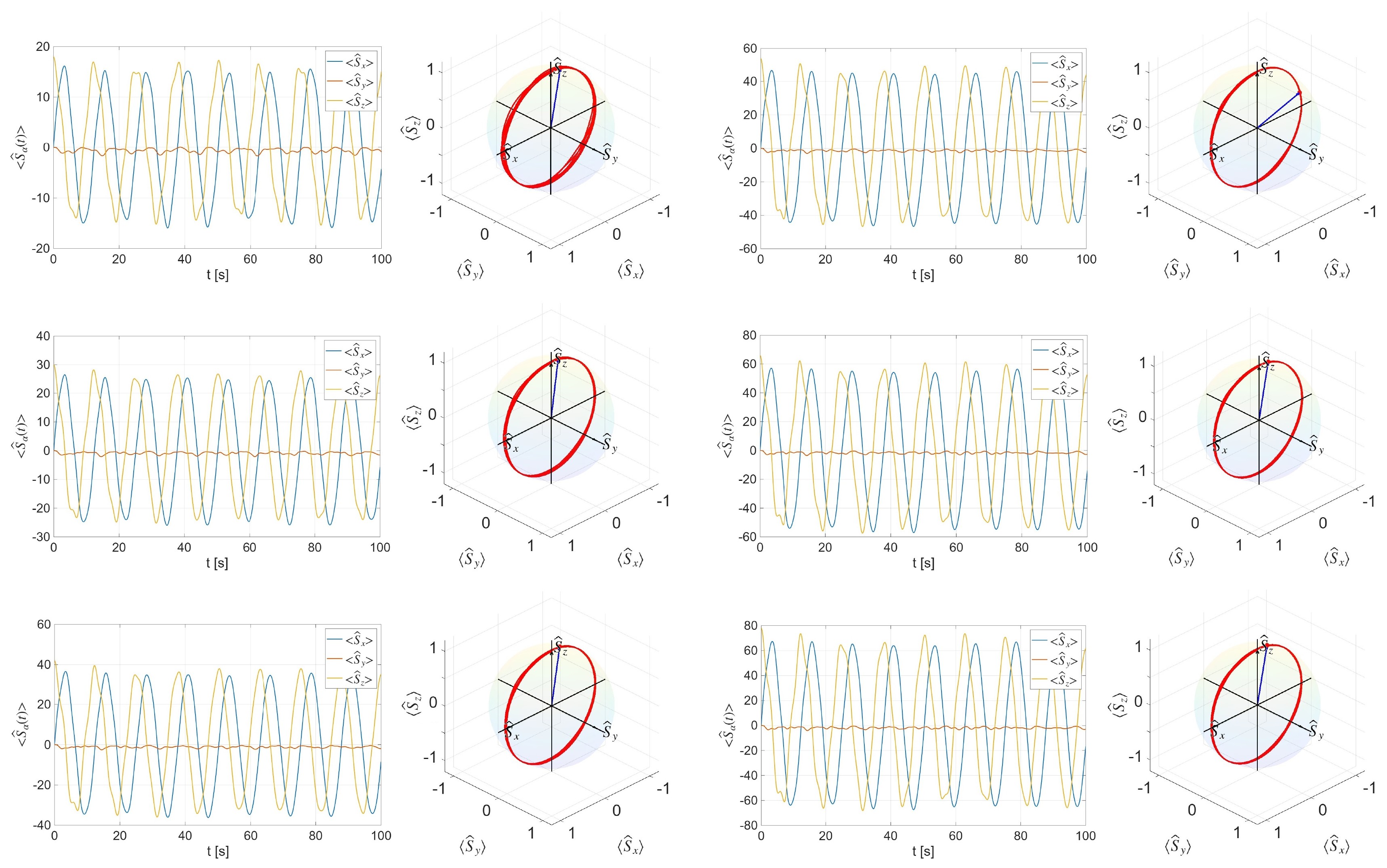}
 \caption{Rotating frame evolution of the spin expectation values over 100 s for a three-spin system with periodic boundary conditions and the corresponding total magnetization on the Bloch sphere after 100 s, for different Floquet spaces, when $B_{0}=1$, $B_{1}=0.5$, $J=1$, and $DMI=1$. The Fourier space range corresponds to different truncation orders, having the Floquet index m = 1 (a), 2 (b), 3 (c), 4 (d), 5 (e), and 6 (f)}
\label{fig:A2}
\end{figure*}

\subsection{Approximate VS full Floquet theory on three spin systems}

Figures~A3 and A4 illustrate the combined influence of energy scales and boundary conditions on the validity of the first-order Floquet Magnus approximation. For the three-spin chain with open boundary conditions (Fig.~A3), pronounced differences from the full Floquet dynamics persist, whereas periodic closure (Fig.~A4) leads to substantially better agreement and more regular collective motion. These results show that the accuracy of the approximate description depends not only on the relative interaction and driving scales but also on the system topology, making the full Floquet treatment particularly important for finite open chains and strong-driving regimes.

\begin{figure*}[htbp]
\centering
\includegraphics[width=1\textwidth]{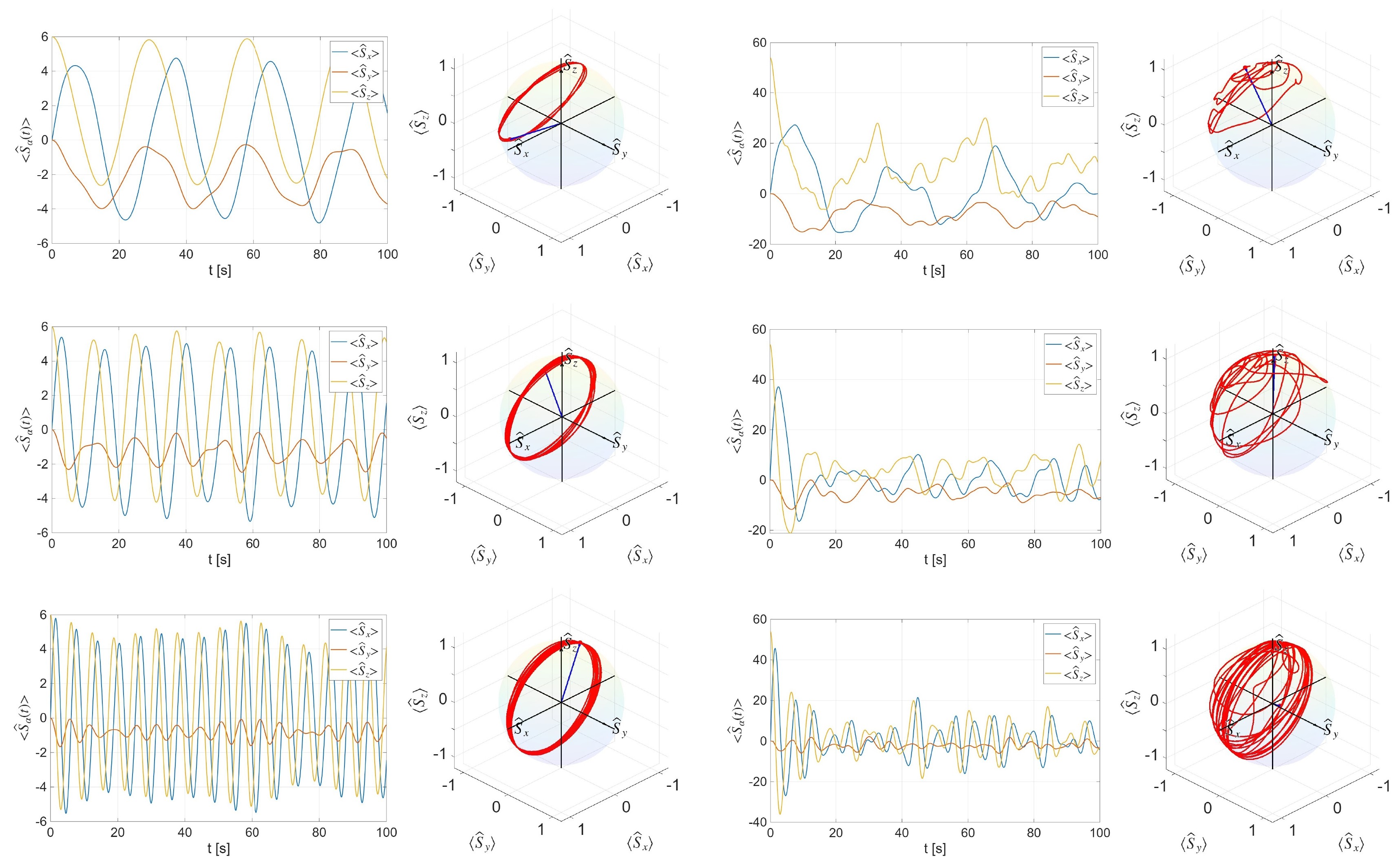}
\caption{Rotating frame evolution of the expectation values over 100 s for a three-
spin system with open boundary conditions and the corresponding total magnetization on the Bloch sphere after
100 s, obtained using Floquet engineering (a-c), and full Floquet theory (d-f), when
B0 = J = D = 1, and B1 = 0.2 (a,d), 0.5 (b,e), and 1 (c,f)}
\label{fig:A3}
\end{figure*}

\begin{figure*}[htbp]
\centering
\includegraphics[width=1\textwidth]{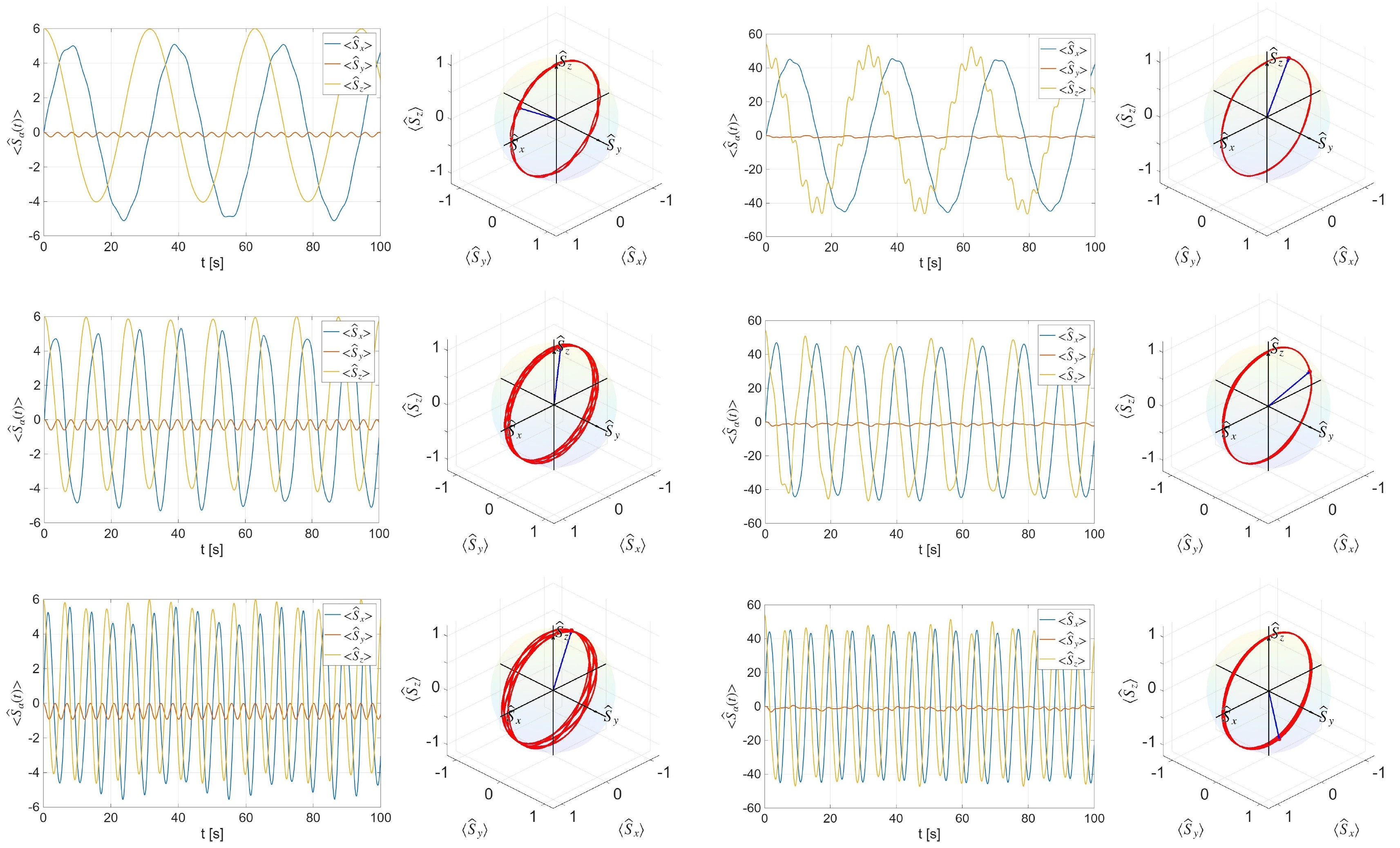}
\caption{Rotating frame evolution of the expectation values over 100 s for a three-
spin system with periodic boundary conditions and the corresponding total magnetization on the Bloch sphere after
100 s, obtained using Floquet engineering (a-c), and full Floquet theory (d-f), when
B0 = J = D = 1, and B1 = 0.2 (a,d), 0.5 (b,e), and 1 (c,f)}
\label{fig:A4}
\end{figure*}

\subsection{Non interacting limit}

Figure $A5$ shows the driven dynamics of a non-interacting three-spin system with open boundary conditions for three transverse driving amplitudes. The spin expectation values exhibit regular coherent oscillations governed by the external driving field. The Bloch-sphere trajectories confirm simple rotation-like motion, consistent with independent spin precession in the absence of spin–spin interactions. Figure $A6$ displays the non-interacting limit for periodic boundary conditions. The dynamics remains purely coherent and regular across all driving amplitudes. The equivalence of open and periodic geometries in this regime confirms that boundary effects arise solely from interactions rather than from the driving fields themselves.

\begin{figure*}[htbp]
\centering
\includegraphics[width=1\textwidth]{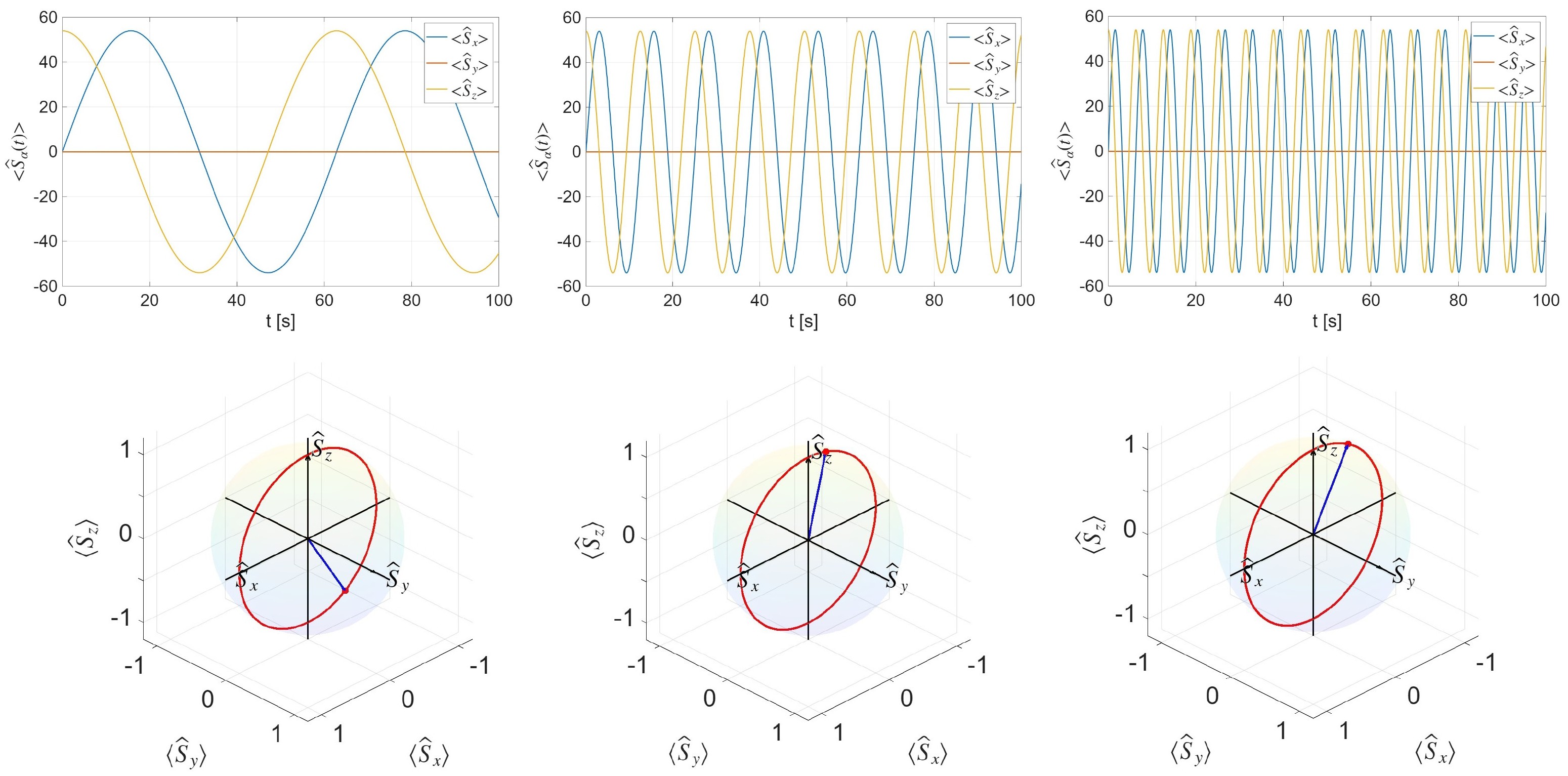}
\caption{Rotating frame evolution of the spin expectation values over 100 s for a three-spin system with open boundary conditions under three driving amplitudes $[B_0, B_1] = [1, 0.1]$ (a), $[1, 0.5]$ (b), and $[1, 1]$ (c), together with the corresponding Bloch-sphere representation of the total magnetization after 100 s}
\label{fig:A5}
\end{figure*}

\begin{figure*}[htbp]
\centering
\includegraphics[width=1\textwidth]{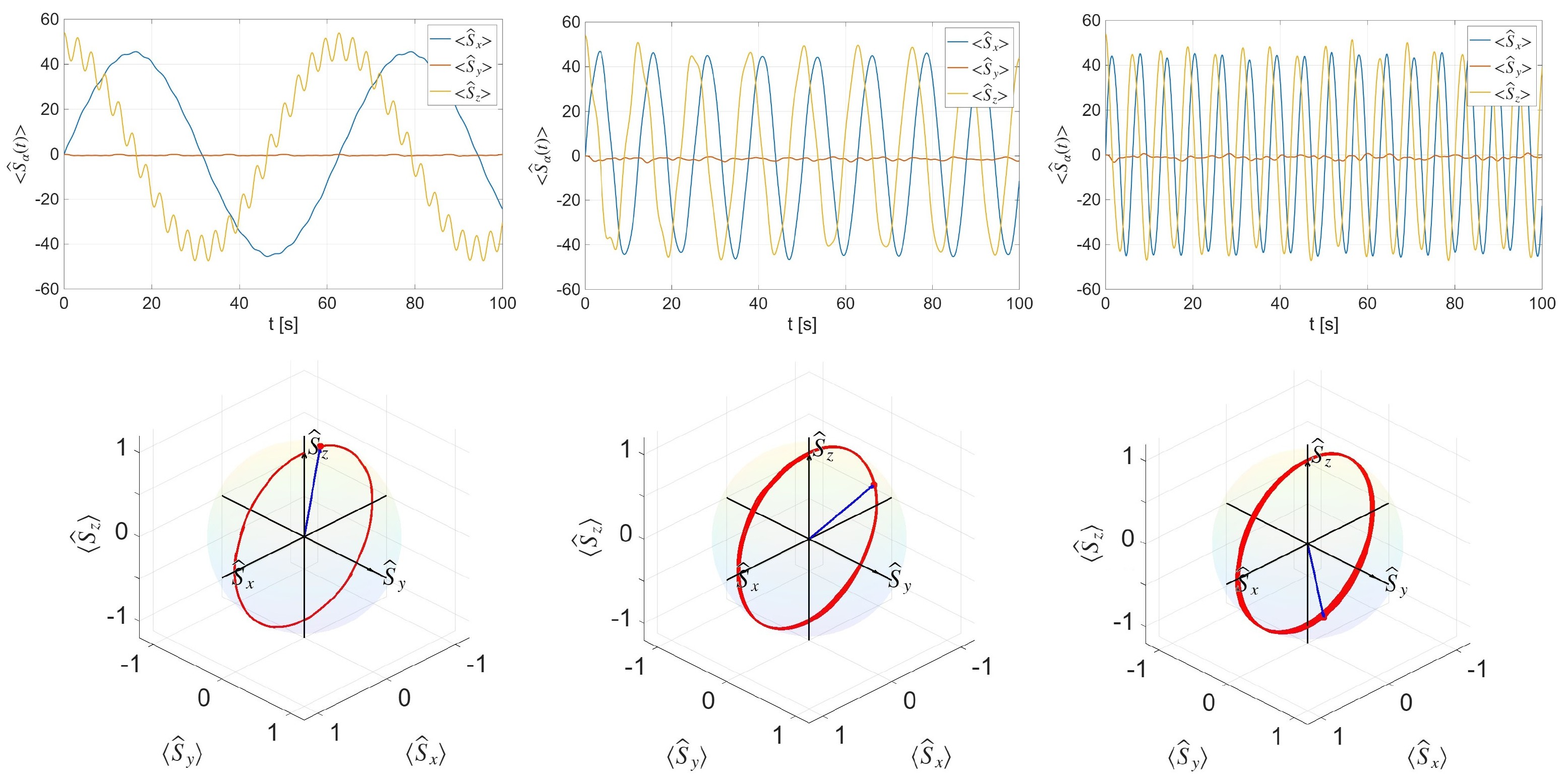}
\caption{Rotating frame evolution of the spin expectation values over 100 s for a three-spin system with periodic boundary conditions under three driving amplitudes $[B_0, B_1] = [1, 0.1]$ (a), $[1, 0.5]$ (b), and $[1, 1]$ (c), together with the corresponding Bloch-sphere representation of the total magnetization after 100 s}
\label{fig:A6}
\end{figure*}

\subsection{The effect of the Dzyaloshinskii–Moriya interaction in the absence of exchange coupling}

Figure $A7$ highlights the effect of the Dzyaloshinskii–Moriya interaction in a two-spin system with exchange coupling neglected. Increasing DMI strength progressively modifies the spin dynamics, generating amplitude modulation and deviations from simple coherent rotations. The Bloch-sphere trajectories reveal the emergence of chiral interaction-induced distortions and multi-frequency behavior.

\begin{figure*}[htbp]
\centering
\includegraphics[width=1\textwidth]{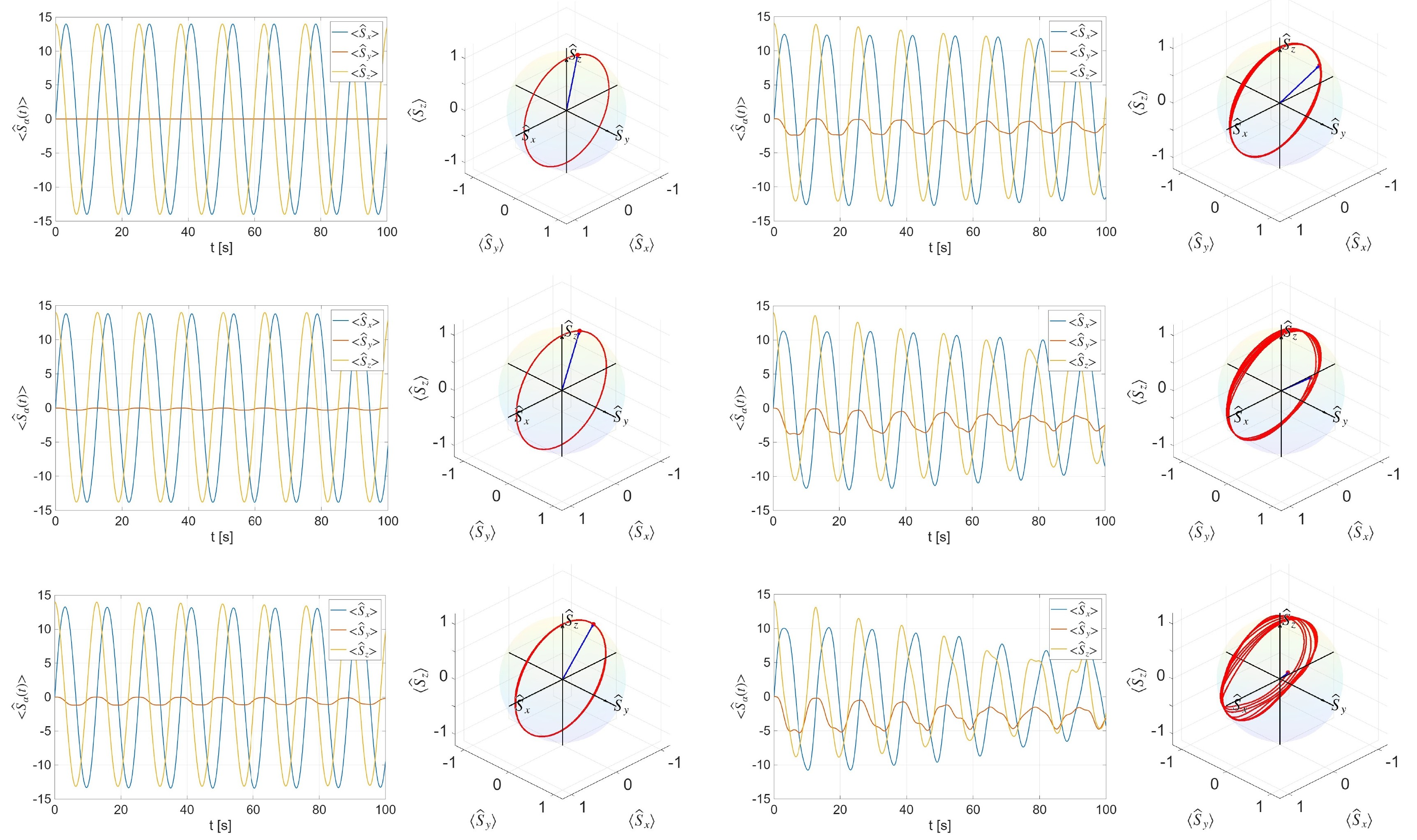}
 \caption{Rotating frame evolution of the spin expectation values over 100 s for a two-spin system, for $[B_0, B_1] = [1, 0.5]$, with the exchange interaction neglected. The Dzyaloshinskii--Moriya interaction is set to 0 (a), 0.2 (b), 0.4 (c), 0.6 (d), 0.8 (e), and 1 (f), together with the corresponding Bloch-sphere representation of the total 
magnetization after 100 s.}
\label{fig:A7}
\end{figure*}

\subsection{Realistic picture: combine effect of exchange coupling and Dzyaloshinskii–Moriya interaction}

Figure $A8$ illustrates the combined influence of exchange coupling and Dzyaloshinskii–Moriya interaction in a two-spin system. The competition between symmetric and antisymmetric interactions produces increasingly complex magnetization trajectories. The resulting dynamics reflect the interplay between collective alignment tendencies and chiral correlation effects.

\begin{figure*}[htbp]
\centering
\includegraphics[width=1\textwidth]{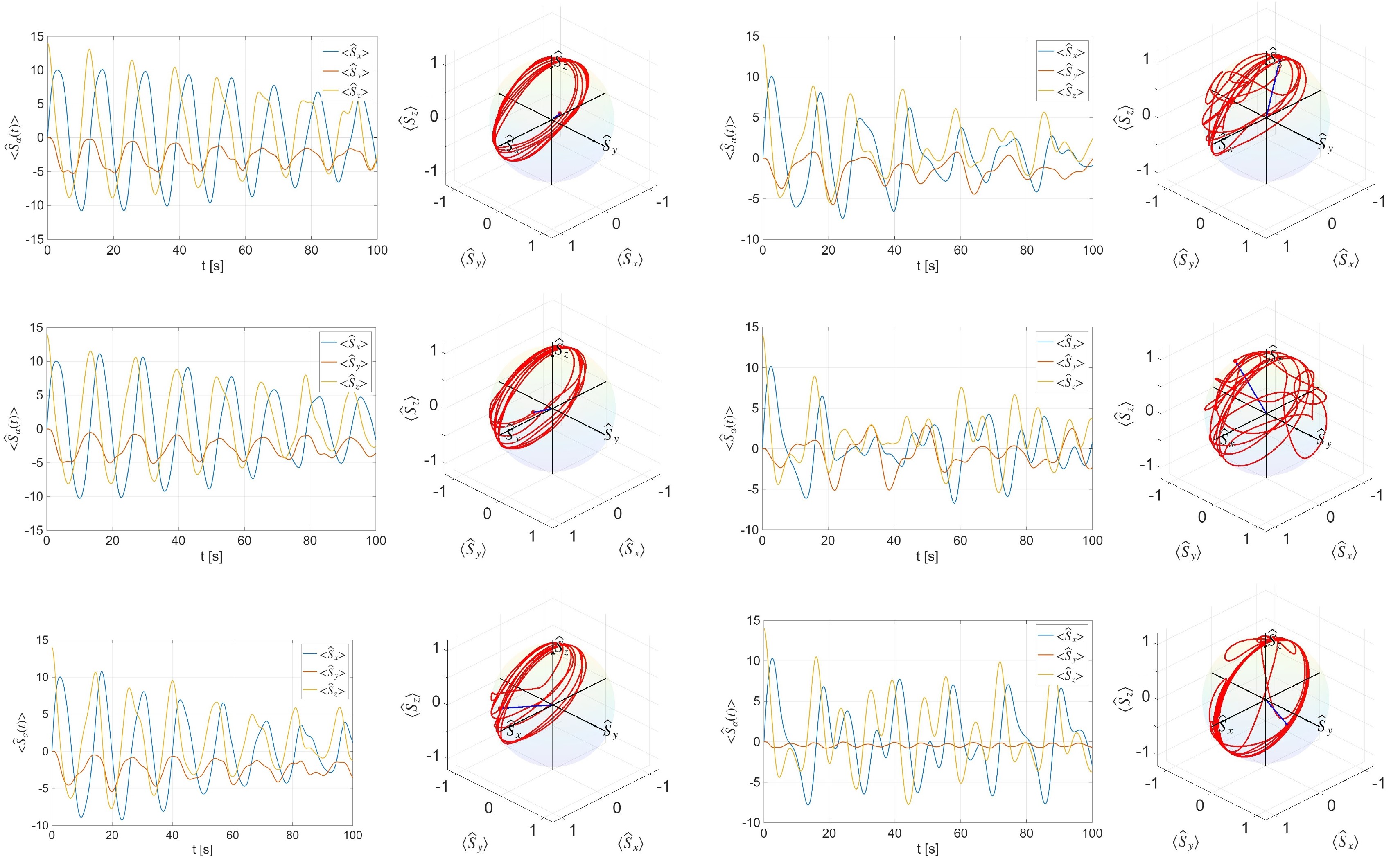}
 \caption{Rotating frame evolution of the spin expectation values over 100 s for a two-spin system at $[B_0, B_1] = [1, 0.5]$, with the Dzyaloshinskii--Moriya interaction fixed to 1. The exchange coupling strength is varied as 0 (a), 0.2 (b), 0.4 (c), 0.6 (d), 0.8 (e), and 1 (f), together with the corresponding Bloch-sphere trajectories of the total magnetization after 100 s.}
\label{fig:A8}
\end{figure*}

\section{Predicted Floquet trajectories and DMI-induced distortions to measurable observables in realistic platforms}

The Floquet trajectories on the Bloch sphere presented in this work are compatible with the coherent implementation of single-qubit logic operations (Pauli X, Y, Z, and Hadamard gates). The specific dynamics analyzed here correspond to a Pauli-Y gate, whose temporal evolution and gate fidelity are shown to be strongly influenced by the Dzyaloshinskii-Moriya interaction (DMI). The extreme low-dimensional limit investigated in this study provides a unique opportunity to uncover the fundamental mechanisms governing spin dynamics, clearly revealing the role of boundary conditions and DMI-induced decoherence. Such effects are often masked in extended spin chains or large two-dimensional lattices, where collective excitations and finite-size averaging dominate the dynamics. Moreover, the complete Floquet description becomes essential when the driving amplitude is comparable to the characteristic interaction energies (exchange and DMI), or when the boundaries are open, allowing an accurate description of the quantum evolution and gate fidelity encoded in spin expectation values (e.g., $\langle \hat{S}_z \rangle$ for Pauli-X or Pauli-Y operations).

Beyond one-dimensional systems, the theoretical framework developed here constitutes a first step toward Floquet engineering of two-dimensional magnetic lattices. In recent work based on exact diagonalization \cite{Tiusan2025}, we demonstrated that the Dzyaloshinskii-Moriya interaction stabilizes chiral skyrmionic states, giving rise to a quantum skyrmion phase under periodic boundary conditions (PBC), whereas open boundary conditions (OBC) favor classical-like, topologically protected skyrmions. Single-qubit quantum gates (Pauli X, Y, Z, and Hadamard) can be implemented on both types of skyrmionic qubits through magnetic-field control and resonant photonic driving. That study further revealed the dual role of the DMI: while it is indispensable for stabilizing skyrmionic qubits, it simultaneously constitutes an intrinsic source of decoherence during gate operations. The present one-dimensional results provide direct microscopic confirmation of this fundamental competition between topological stabilization and coherent quantum control.

Ultimately, translating these theoretical concepts into practical quantum hardware will require the realization of atomically precise low-dimensional magnetic materials, including molecular spin chains \cite{KK} and two-dimensional van der Waals ferrimagnetic magnets \cite{ZL, HL}, which constitute promising platforms for quantum spintronics. Such systems offer the possibility of engineering both the exchange interaction (\textit{J}) and the Dzyaloshinskii-Moriya interaction (\textit{D}) over a broad parameter range, thereby reducing thermal and structural noise while enhancing macroscopic quantum coherence. Through appropriate material selection, interface engineering, and growth conditions, the relative magnitudes of \textit{J} and \textit{D} can be optimized to maximize gate fidelity and coherence times. From a theoretical perspective, the validity of approximate Floquet-engineering approaches versus a full Floquet treatment is ultimately determined by the ratio between the driving amplitude and the competing intrinsic interaction energies, primarily \textit{J} and \textit{D}, and by the boundary conditions. Consequently, accurate modeling of realistic quantum spintronic devices requires a full Floquet description whenever these energy scales become comparable.

\end{appendix}

\clearpage

%\section*{References}

\bibliography{apssamp}

\end{document}